\documentclass[aps, prx, reprint,footinbib,floatfix,superscriptaddress]{revtex4-2}
\usepackage{graphicx}
\usepackage{indentfirst}
\usepackage{braket}
\usepackage{float}
\usepackage{amsmath}
\usepackage{physics}
\usepackage{amssymb}
\usepackage{CJK}
\usepackage{esint}
\usepackage{color}
\usepackage[T1]{fontenc}
\usepackage{amsfonts}
\usepackage{footmisc}
\usepackage{scrextend}
\usepackage{multirow}
\usepackage[outdir=./]{epstopdf}
\usepackage{xcolor}
\usepackage[english]{babel}
\usepackage{url}
\usepackage{comment}
\usepackage{array}
\usepackage{subfiles}
\usepackage{tikz}
\usetikzlibrary{shapes,arrows}
\usetikzlibrary{positioning,calc}
\usepackage{mathrsfs}
\usepackage{mathtools}
\usepackage[mathscr]{eucal}
\usepackage{multirow}
 \usepackage{booktabs} 

\usepackage[hyperfootnotes=false]{hyperref}
\usepackage{cleveref}
\definecolor{darkblue}{rgb}{0,0,0.5}
\hypersetup{
colorlinks=true,
linkcolor=black,
filecolor=blue,
citecolor=darkblue,  
urlcolor=cyan,
}

\newtheorem{theorem}{Theorem}

\usepackage{trimclip}

\makeatletter
\DeclareRobustCommand{\shortto}{%
  \mathrel{\mathpalette\short@to\relax}%
}

\newcommand{\short@to}[2]{%
  \mkern2mu
  \clipbox{{.5\width} 0 0 0}{$\m@th#1\vphantom{+}{\shortrightarrow}$}%
  }
\makeatother

\usepackage{pict2e,picture,graphicx}

\makeatletter
\DeclareRobustCommand{\Arrow}[1][]{%
\check@mathfonts
\if\relax\detokenize{#1}\relax
\settowidth{\dimen@}{$\m@th\rightarrow$}%
\else
\setlength{\dimen@}{#1}%
\fi
\sbox\z@{\usefont{U}{lasy}{m}{n}\symbol{41}}%
\begin{picture}(\dimen@,\ht\z@)
\roundcap
\put(\dimexpr\dimen@-.7\wd\z@,0){\usebox\z@}
\put(0,\fontdimen22\textfont2){\line(1,0){\dimen@}}
\end{picture}%
}
\makeatother

\def\be{\begin{equation}}
\def\ee{\end{equation}}
\def\ba{\begin{eqnarray}}
\def\ea{\end{eqnarray}}
\def\bal{\begin{equation}\begin{aligned}}
\def\eal{\end{aligned}\end{equation}}

\def\d{^\dagger}
\def\bp{\begin{pmatrix}}
\def\ep{\end{pmatrix}}

\usepackage{bm}

\newcommand{\calA}{{\cal A}}

\newcommand{\calB}{{\cal B}}

\newcommand{\calD}{{\cal D}}

\newcommand{\calL}{{\cal L}}

\newcommand{\calN}{{\cal N}}

\newcommand{\calQ}{{\cal Q}}

\newcommand{\calV}{{\cal V}}
\newcommand{\calW}{{\cal W}}
\newcommand{\calX}{{\cal X}}

\newcommand{\1}{^{(1)}}

\newcommand{\QZ}[1]{{{\textcolor{black}{#1}}}}

\usepackage[normalem]{ulem}
\usepackage{xr}
\begin{document}

\title{Distributed dual-comb spectroscopy}

\author{Haowei Shi}
\thanks{These two authors contributed equally.}
\address{
Ming Hsieh Department of Electrical and Computer Engineering, University of Southern California, Los
Angeles, California 90089, USA
}

\author{Kartikeya Arora}
\thanks{These two authors contributed equally.}
\address{
Ming Hsieh Department of Electrical and Computer Engineering, University of Southern California, Los
Angeles, California 90089, USA
}

\author{Zheshen Zhang}
\address{Department of Electrical Engineering and Computer Science, University of Michigan, Ann Arbor, MI 48109, USA}

\author{Quntao Zhuang}
\email{qzhuang@usc.edu}

\address{
Ming Hsieh Department of Electrical and Computer Engineering, University of Southern California, Los
Angeles, California 90089, USA
}
\address{
Department of Physics and Astronomy, University of Southern California, Los
Angeles, California 90089, USA
}
\begin{abstract}
We propose \emph{distributed dual-comb spectroscopy}, a quantum-enhanced sensing architecture that simultaneously exploits spatial- and spectral-domain entanglement to improve the extraction of global spectroscopic information across broadband frequency ranges. 
Our scheme consists of a network of Mach--Zehnder interferometers interrogating spatially separated samples, each driven by a classical frequency comb at the bright port. 
A single two-mode squeezed frequency comb is distributed through a passive beam-splitter array into $K$ entangled combs, establishing multipartite correlations across sensors while preserving intrinsic spectral entanglement. The protocol is compatible with both heterodyne detection and division detection, and enables parallel estimation of weighted-average absorption features over multiple frequency components. 
Compared with spatially separable quantum strategies requiring $K$ independently engineered nonclassical comb sources, our approach achieves enhanced signal-to-noise performance using only a single quantum resource distributed across the network. 
These results establish a scalable route toward broadband distributed quantum spectroscopy with simultaneous spatial and spectral advantage.


\end{abstract}
\date{\today}
\maketitle

\section{Introduction}

Real-world sensing tasks are inherently complex, often involving both spatially distributed parameters and broadband spectral features. 
In the spatial domain, the quantities of interest may be spread across multiple locations, necessitating sensor networks capable of extracting global information from distributed measurements. 
Distributed quantum sensing~\cite{zhang2021distributed,zhuang2018distributed,ge2018distributed} addresses this challenge by exploiting multipartite entanglement across spatially separated probes, enabling sensitivity enhancements beyond classical limits. 
Such techniques have been successfully applied to radiofrequency detection~\cite{xia2020demonstration}, photonic phase estimation~\cite{guo2020distributed}, and force sensing~\cite{xia2023entanglement}.

In parallel, many sensing applications---spectroscopy in particular---are characterized by a large range of relevant frequencies. 
Dual-comb spectroscopy (DCS)~\cite{Picque2019, Coddington:16, Fortier2019,Martin-Mateos_imagingOptica, Vicentini2021,Coddington2009RapidAP, kippenberg_ranging2011, Lukashchuk2022, Muh_ranging2018, Caldwell2022}, based on optical frequency combs~\cite{hansch2006,hall2006} spanning broad bandwidths with ultra-precise mode spacing, provides a powerful platform for the simultaneous acquisition of absorption properties over wide spectral ranges. 
Quantum-enhanced DCS~\cite{shi2023entanglement,herman2024squeezed,hariri2025entangled,shi2025theory} further improves performance by utilizing spectral-domain squeezing and entanglement to enhance the signal-to-noise ratio (SNR) in broadband measurements.

In this work, we propose \emph{distributed dual-comb spectroscopy}, a quantum sensing architecture that combines the advantages of both spatial- and spectral-domain entanglement to enhance the extraction of global information across a large frequency range simultaneously. 
Our sensor network consists of a set of Mach--Zehnder interferometers (MZIs), each receiving a classical frequency comb through its bright port. 
A single two-mode squeezed frequency comb is distributed via a passive beam-splitter array into $K$ entangled combs, thereby establishing correlations across the spatial domain (through mode mixing) and across the spectral domain (through two-mode squeezing). 
These quantum combs are injected into the dark ports of the $K$ interferometers.

To read out the sensing signal, we employ local-oscillator (LO) combs to enable heterodyne~\cite{hariri2025entangled} \QZ{or a division receiver~\cite{herman2024squeezed}} to detect the interfered outputs, enabling the simultaneous estimation of weighted-average absorption features from $K$ samples over multiple frequency components.
Notably, even at the single-sample level, the balanced MZI configuration mitigates the low-brightness limitation of previously proposed entangled frequency-comb schemes~\cite{hariri2025entangled}. 
Moreover, achieving comparable performance using spatially separable strategies would require $K$ independently engineered quantum frequency combs, resulting in a substantial resource overhead. 
In contrast, our approach relies on only a single quantum-engineered comb distributed across the entire heterodyne sensor network, providing a scalable route toward broadband quantum-enhanced distributed spectroscopy.



\begin{figure}
    \centering
    \includegraphics[width=\linewidth]{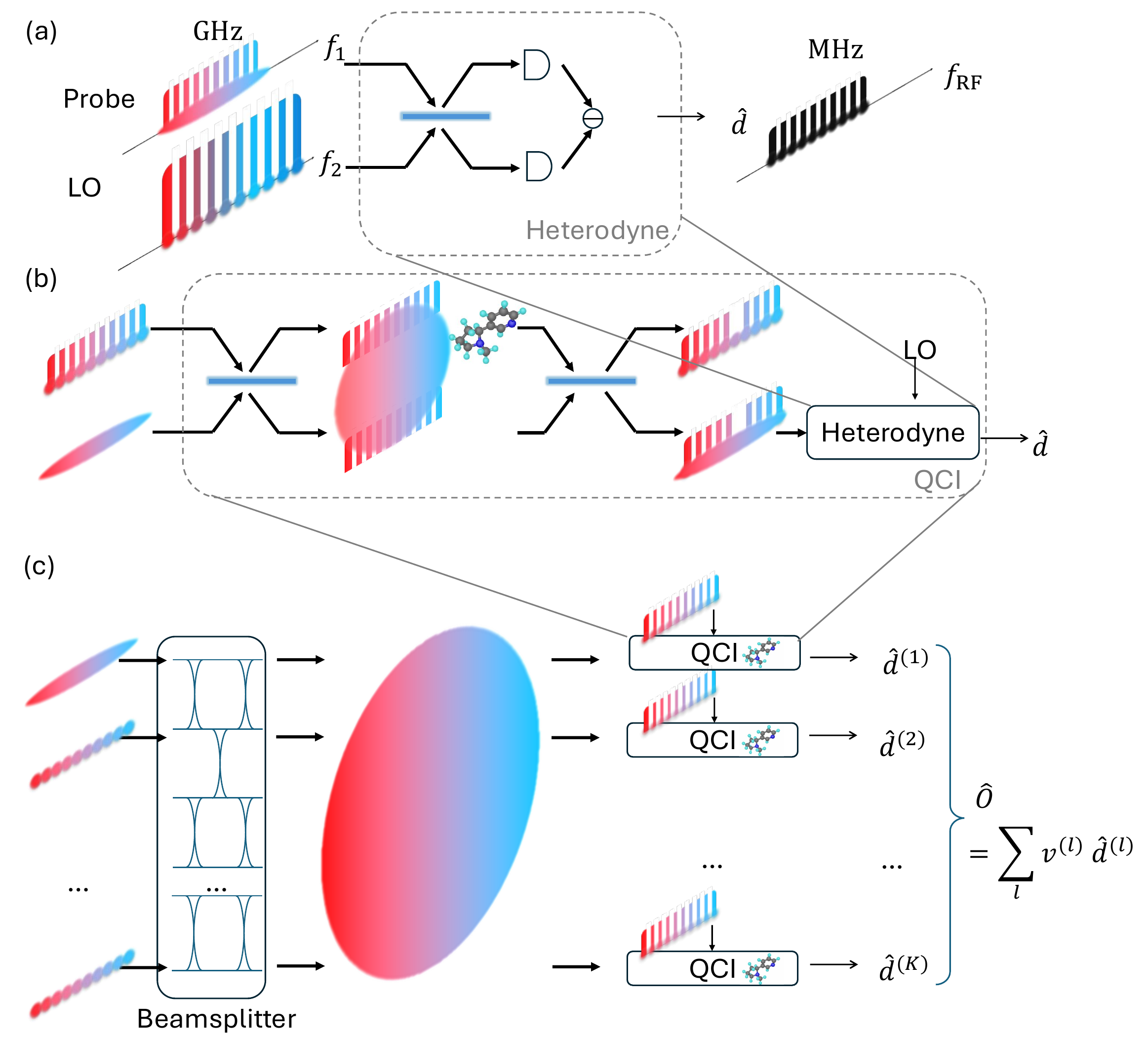}
    \caption{(a) Dual comb spectroscopy \QZ{with a heterodyne receiver}; (b) Quantum comb interferometry (QCI) \QZ{based on a Mach-Zehnder interferometer}; (c) Distributed QCI for $K$ samples, where the $K$ inputs are entangled via a beamsplitter \QZ{at input sides $A$ and $B$, respectively}.}
    \label{fig:schematic_concept}
\end{figure}

\section{ Quantum frequency combs and dual comb spectroscopy}
In dual comb spectroscopy (see Fig.~\ref{fig:schematic_concept}a), two combs at slightly different repetition rates $\omega_r/2\pi$ and $(\omega_r+\Delta \omega_r)/2\pi$ are utilized. The signal comb
\bal 
\hat A(t)&=\frac{1}{\sqrt{T}}\sum_{n=-N}^N\sum_{m=-2N}^{2N} e^{-i (n\omega_r+m\Delta\omega_r) t} (A_n\delta_{n,m} + \hat \calA_{n,m})
\label{eq:combA}
\eal 
has non-zero amplitudes $A_n$ at frequencies $n(\omega_r+\Delta \omega_r)/2\pi$~\cite{shi2023entanglement,shi2025theory}. We assume that the acquisition time $T$ is long enough to resolve the frequency of interest. Here $\hat \calA_{n,m}$'s are canonical annihilation operators that describe the quantum fluctuation properties. While generally the LO comb may also be quantum engineered, in this work we focus on \QZ{the heterodyne case with a strong LO}~\cite{shi2025theory}, \QZ{where the LO can} be effectively described semi-classically as
\be 
\hat{B}(t)=\frac{1}{\sqrt{T}}\sum_{n=-N}^N  e^{-i n\omega_r t}  B_n.
\label{eq:combB,classical}
\ee 



In heterodyne DCS, the signal comb is used to probe the sample, while the LO comb interferes with the signal comb to produce the beat notes at frequencies $m\Delta\omega_r$ and directly measures an ultra broadband spectrum via a spectral compression. We will focus on heterodyne detection in the main text, while \QZ{the division receiver version is given} in Methods in parallel.

In linear absorption spectroscopy, the quantum model of a sample is an array of bosonic loss channels~\cite{weedbrook2012gaussian} with a frequency-dependent complex-valued transmissivity spectrum $\tau(\omega)=\sqrt{\kappa(\omega)}e^{i\phi(\omega)}$, which combines frequency-dependent power transmissivity $\kappa(\omega)$ and phase shift $\phi(\omega)$. 
When the frequency comb goes through the channel, the output $\calL[\hat{A}(t)]$ \QZ{has the same form as} Eq.~\eqref{eq:combA} with
\bal 
\centering
\calL: &A_n&&\to A_n^\prime= \tau_n A_n \,,\\
&\hat \calA_{n,m}&&\to \hat \calA^\prime_{n,m}= \tau_n\hat \calA_{n,m}+\sqrt{1-|\tau_n|^2}\hat \calV_{n,m}
\label{loss_channel}
\eal 
with $\hat \calV_{n,m}$ denoting the environment mode due to loss.  For simplicity, we consider $\hat \calV_{n,m}$ to be vacuum and $\tau(\omega)$ to be uniform within the neighborhood of each comb, $\tau(n\omega_r+m \Delta \omega_r)\equiv \tau_n$, for all relevant $-2N\le m\le 2N$. 

To enable our discussion on the distributed DCS, we briefly review the results about dual-comb interferometry for a single sample.

Given a signal comb $\hat{A}(t)$ \QZ{and an LO comb} $\hat{B}(t)$, the quantum operator describing the heterodyne in the spectral domain is
\be 
\hat{D}_m\equiv \int e^{im \Delta\omega_r t}\left\{\hat A(t)^\dagger \hat B(t)+\hat B^\dagger(t)\hat A(t)\right\}.
\label{difference_current_main}
\ee 
Keeping the low frequency terms, combining Eqs.~\eqref{eq:combA} and \eqref{eq:combB,classical},  
\QZ{and taking a uniform real LO, $B_m=B$, we find that the rescaled quantum operator $\hat{d}_m=\hat{D}_m/B$ has} the mean and noise terms
\begin{align} 
&\expval{\hat{d}_m}= A_m + A_{-m}^\star.
\label{d_m_het2}
\\
&
\Delta \hat{d}_m=
\sum_{n=-N}^N (\hat \calA_{n,m}+\hat \calA_{n,-m}^{\dagger}).
 \label{noise_maintext2}
\end{align}
We see that \QZ{heterodyne using a uniform LO comb with different spacing from the signal comb} is going to extract all mean values of the signal comb, with added noise. To eliminate the degeneracy between two frequencies of $A_m$ and $A_{-m}^\star$ in Eq.~\eqref{d_m_het2}, several approaches have \QZ{been developed~\cite{hariri2025entangled}; a simple approach} is to let $A_{m<0}=0$.


\section{Single-sensor case: resolving the low-brightness issue}

From Eq.~\eqref{noise_maintext2}, in order to directly enjoy quantum advantage from DCS, one can perform quantum engineering of the mode pairs $\hat \calA_{n,m}+\hat \calA_{n,-m}$ $\hat \calA_{n,m}+\hat \calA_{-n,-m}$~\cite{shi2025theory,shi2023entanglement}. By engineering the two-mode squeezing between pairs of $(\hat \calA_{n,m}, \hat \calA_{n,-m})$ or $(\hat \calA_{n,m}, \hat \calA_{-n,-m})$, the noise can be therefore suppressed; at the same time, we want to \QZ{maintain a large signal} in the mean values of Eq.~\eqref{d_m_het2}.   
Therefore, quantum-enhanced heterodyne DCS relies on the engineering of a bright quantum comb with non-zero mean $A_m$'s as strong as classical benchmarks and suitable two-mode squeezing on the proper sidebands. 

Ref.~\cite{hariri2025entangled} engineers such a comb by displacing a zero-mean entangled comb \QZ{with a classical comb; however,} that \QZ{leads to a waste of} the classical comb power due to the necessarily almost transparent beamsplitter preserving most of the entangled comb while only mixing in a small portion of the classical comb. In this approach, while the noise is suppressed, the signal is weaker due to small amplitudes $A_m$. While quantum advantage can be found at the same power level, \QZ{this approach drastically limits} the operating power regime of quantum DCS.

As depicted in Fig.~\ref{fig:schematic_concept}b, we propose an interferometry-based approach to resolve the low-brightness issue of Ref.~\cite{hariri2025entangled}.
Consider a field with quantum engineered noise $\hat{A}^Q (t)\propto \sum_{n,m} \hat \calA_{n,m}^{Q} e^{-i (n\omega_r+m\Delta\omega_r) t}  /\sqrt{T}$ \QZ{with a zero mean field}, and another field $\hat{A}^C (t)$ \QZ{with a strong nonzero mean} $\expval{\hat{A}^C (t)}={A}^C (t)=\sum_{n} A_ne^{-i n(\omega_r+\Delta\omega_r) t} /\sqrt{T}$. The first beamsplitter interferes the zero-mean quantum comb and the strong classical comb to generate
\be 
\hat{A}_\pm (t) =\frac{\hat{A}^Q (t)\pm \hat{A}^C (t)}{\sqrt{2}}.
\label{Apm}
\ee 
The upper arm $\hat{A}_+ (t)$ goes through a sample, described by a quantum channel $\calL$ in Eq.~\eqref{loss_channel}, leading to the output
$
\calL[\hat{A}_+ (t)].
$
At the reverse beamsplitter, we have the dark port output
\begin{align}
\hat{A}^{Q\prime}(t)
&=\frac{1}{\sqrt{2}}\left(\calL[\hat{A}_+ (t)]+\hat{A}_-(t)\right)
\\
&=\hat{A}^Q (t)+\frac{\calL[\hat{A}_+ (t)]-\hat{A}_+ (t)}{\sqrt{2}}.
\label{eq:AQ_prime}
\end{align} 
Combining Eq.~\eqref{Apm} with the above, we find that $1/4$ of the initial classical comb energy effectively picks up the signal, instead of the vanishing portion in Ref.~\cite{hariri2025entangled}. More importantly, the quantum noise is determined by the zero-mean quantum engineered comb $\hat{A}^Q (t)$.

We consider the weak absorption limit from now on for simplicity of description, although this is not required for the protocol to provide quantum advantage. 
\QZ{Consider $\hat{A}^C (t)$ with a strong non-zero mean, and assume that the transmissivities $\tau_n$ are all close to unity}, \QZ{so the perturbation from the sample in the noise is weak:} $\calL[\hat{A}_+ (t)]\simeq \expval{\calL[\hat{A}_+ (t)]} + \hat{A}_+ (t)- \expval{\hat{A}_+ (t)}$. Therefore,
the noise in Eq.~\eqref{eq:AQ_prime} is dominated by the zero-mean comb $\hat{A}^Q (t)$, while the mean is dominated by the classical term in $(\calL[\hat{A}_+ (t)]-\hat{A}_+ (t))/{\sqrt{2}}$.

We can therefore \QZ{perform DCS} between the interfered output comb $\hat{A}^{Q\prime}(t)$ and a strong LO comb with repetition rate $\omega_r/2\pi$ \QZ{to resolve all means and variances} from Eqs.~\eqref{d_m_het2} and \eqref{noise_maintext2}
\begin{align} 
&\expval{\hat{d}_m}= \frac{1}{2}A_m(\tau_m-1).
\label{d_m_het2_final}
\\
&
\Delta \hat{d}_m=
\sum_{n=-N}^N (\hat \calA_{n,m}^{Q}+\hat \calA_{n,-m}^{Q\dagger}).
 \label{noise_maintext2_final}
\end{align}
Note that here we have taken $A_{m\le0}=0$ to eliminate degeneracy for simplicity. Here, $\hat \calA_{n,m}^{Q}$'s are the noise terms of the zero-mean quantum comb $\hat{A}^Q (t)$ in the mode definition of Eq.~\eqref{eq:combA}\QZ{, while $A_m$'s are the classical mean fields} of the strong classical comb $\hat{A}^C (t)$.

Therefore, we can suppress noise by the zero-mean quantum engineered comb. \QZ{Compared with the previous protocol}~\cite{hariri2025entangled}, \QZ{here the amplitude SNR, up to a factor of one half, enjoys} the entire classical comb power, while in \cite{hariri2025entangled} most of the classical comb is wasted due to \QZ{an almost transparent beamsplitter}.

It is worth mentioning that parallel to the heterodyne approach, one can also adopt a division receiver, where one interferes \QZ{the LO comb and the signal comb to probe the samples}\QZ{. While the differences are studied} in Ref.~\cite{shi2025theory}: \QZ{squeezing is self-referenced for a division receiver} while cross-referenced for heterodyne and the two approaches possess \QZ{different sensitivities to loss, we emphasize} that in this weak absorption regime, given input combs with identical noise covariances, up to the frequency
shift between squeezing referenced to the signal lines and to the LO
lines, the heterodyne and division readouts have identical SNR up to a
constant set by the comb amplitudes \QZ{(see Methods and Appendix}~\ref{Equivalence_Appendix}). This equivalence provides the basis of our unified treatment in designing the distributed DCS setup as we detail below.




\section{Distributed dual-comb spectroscopy}
As shown in Fig.~\ref{fig:schematic_concept}c, we consider $K$ different samples, \QZ{each with an absorption spectrum} $\{\tau_m^{(\ell)}\}_{m=-N}^N$, for $1\le \ell\le K$. \QZ{We use the superscript} $(\ell)$ to index the $\ell$-th \QZ{sample in the spatial domain, to distinguish it from} the subscript $m$ which indexes the frequency component. Each sample is probed by a quantum comb and a classical comb \QZ{in an interferometer set-up}, leading to the heterodyne mean and noise analogous to Eqs.~\eqref{d_m_het2_final} and \eqref{noise_maintext2_final}
\begin{align} 
&\expval{\hat{d}_m^{(\ell)}}= \frac{1}{2}A_m^{(\ell)}(\tau_m^{(\ell)}-1).
\label{d_m_het2_final}
\\
&
\Delta \hat{d}_m^{(\ell)}=
\sum_{n=-N}^N (\hat \calA_{n,m}^{Q,(\ell)}+\hat \calA_{n,-m}^{Q,(\ell)\dagger}).
 \label{noise_maintext2_final}
\end{align}

Our goal is to extract global properties of the samples, for example the weighted averages of the absorption coefficients, 
\be 
\theta_m=\sum_{\ell=1}^K w^{(\ell)}(\tau_m^{(\ell)}-1),  \quad -N\le m \le N
\ee 
across all the frequency bins under investigation. Here, we define the average with unity being subtracted for convenience. Such a global parameter can model different application scenarios. For example, if one wants to understand the average absorption for samples in a region, then one can choose the weight $w^{(\ell)}=1/M$\QZ{; if in a certain application}, gas samples across different locations are diffused from a single source point of leakage, and one wants to locate the source point of leakage, then one may take the gradient, e.g. $\bm w=(1,0,-1)$ for three points. We will detail some application scenarios in the next section, after we provide the basic tools for the analyses.

To construct an estimator for the parameter of interest, we take a linear weighted sum of \QZ{the re-scaled heterodyne measurements,}
\be 
\hat{O}_m=\sum_\ell v^{(\ell)} \hat{d}_m^{(\ell)}.
\label{eq:estimator}
\ee 
\QZ{The mean value of the observable is}
\be 
\expval{\hat{O}_m}=\frac{1}{2}\sum_\ell v^{(\ell)} A_m^{(\ell)}(\tau_m^{(\ell)}-1),
\ee 
\QZ{and the noise is}
\be 
\Delta \hat{O}_m=\sum_\ell v^{(\ell)}\sum_{n=-N}^N (\hat \calA_{n,m}^{Q,(\ell)}+\hat \calA_{n,-m}^{Q,(\ell)\dagger}).
\ee 
By choosing the parameters $v^{(\ell)}$ and the classical comb amplitudes $A_m^{(\ell)}$ according to
\be 
v^{(\ell)} A_m^{(\ell)}= 2w^{(\ell)},
\ee 
we can obtain an unbiased estimator $\hat O_m$ with the mean $\expval{\hat{O}_m}=\theta_m$.
Under such an unbiased condition, the noise of the \QZ{observable} is described by the operator
\be 
\Delta \hat{O}_m=2\sum_\ell \frac{w^{(\ell)}}{A_m^{(\ell)}}\sum_{n=-N}^N (\hat \calA_{n,m}^{Q,(\ell)}+\hat \calA_{n,-m}^{Q,(\ell)\dagger}).
\label{eq:varO_unbiased}
\ee

To reduce ${\rm Var}[\Delta \hat{O}_m]$, we need to squeeze $\sum_\ell w^{(\ell)} (\hat \calA_{n,m}^{Q,(\ell)}+\hat \calA_{n,-m}^{Q,(\ell)\dagger})$. Such an array of entangled combs can be generated by passing a single two-mode-squeezed quantum comb $\hat \calA_{n,m}^Q$, with ${\rm var} (\hat \calA_{n,m}^{Q}+\hat \calA_{n,-m}^{Q\dagger})=1/G_n$, through a beamsplitter. We use a beamsplitter to distribute the quantum comb, with beamsplitter weights $\bm q\equiv \{q^{(\ell)}\}_{\ell=1}^K$ such that $\hat \calA_{n,m}^{Q,(\ell)}=q^{(\ell)} \hat \calA_{n,m}^{Q}+{\rm vac}$, where ${\rm vac}$ represents vacuum noises from the other $K-1$ vacuum inputs.
Setting the beamsplitter ratio to $q^{(\ell)}={w^{(\ell)}}/{\sqrt{\sum_{\ell=1}^K |w^{(\ell)}|^2}}$, we have
\be 
{\rm Var}[\Delta \hat{O}_m]=4(\sum_n \frac{1}{G_n})\sum_\ell |\frac{w^{(\ell)}}{A_m^{(\ell)}}|^2.
\label{eq:varO_qBSopt}
\ee

\QZ{Now consider a total power constraint} over the $K$ samples $\sum_{\ell=1}^K |A_m^{(\ell)}|^2\cdot \hbar\Omega_c/T=2P_m$ where $\Omega_c$ \QZ{is the carrier frequency}, $T$ \QZ{is the integration time}, \QZ{and the factor of 2 arises because} the MZI halves the power exposure on each sample. \QZ{Using a Lagrangian multiplier}, we find the optimal power distribution as
$ 
|A_m^{(\ell)}|=\sqrt{{\frac{P_mT}{\hbar\Omega_c}|w^{(\ell)}|}/{\sum_\ell |w^{(\ell)}|}}
$,
leading to the minimum noise
\be 
{\rm Var}[\Delta \hat{O}_m]=\frac{2\hbar\Omega_c}{P_mT}(\sum_n \frac{1}{G_n})(\sum_\ell |w^{(\ell)}|)^2.
\label{eq:varO_opt}
\ee 

\QZ{For a per-sample power constraint}, the result is given as Eq.~\eqref{eq:minvar_uniform}.

It is noteworthy that distribution preserves the receiver equivalence, up to the same mapping between self-referenced and cross-referenced squeezing. For any number of samples and any target
weights, both receivers are optimised by the same beamsplitter,
postprocessing weights and comb amplitudes in the weak sample absorption regime \QZ{(see Methods and Appendix}~\ref{Equivalence_Appendix}).



\section{Application examples}

In this section, we consider applications of the distributed DCS, focusing on gas sensing. We begin with the sample model of transmissivities $\tau$'s in Eq.~\eqref{loss_channel} for gas sensing. For simplicity, we take the sample phase to be negligible, so the transmissivity of the samples $\tau=\sqrt{\kappa}$ is real-valued, accounting for the gradient of the gas density and absorbance. We also assume absorption is weak $\tau\simeq 1$, as we have already assumed in earlier analyses. Given a gas with density $\rho$ and a fixed gas sensor chamber configuration, \QZ{the transmissivity is $\tau=\exp(-\sigma \rho/2)\simeq 1-\sigma \rho/2$ in the weak absorption limit}, where $\sigma$ is the cross-section of absorption. 

DCS is powerful due to its wide frequency coverage, and suitable for sensing of $S\gg1$ different gas species simultaneously. We assume that the gases are distinct in absorption frequencies---each covered by the comb line $m_g$\QZ{---and their absorption peaks are at} $m_g\omega_r$. The distributed DCS further allows \QZ{spatial domain distribution; thus, we consider} $K$ DCS sensors at different locations. Overall, we are \QZ{interested in the estimation} of functions of gas \QZ{densities} $\rho_g^{(\ell)}$, where $\ell\in \sin[1,K]$ is the spatial index and $g\in [1,S]$ is the gas index. The densities correspond to the transmissivities at frequency $m_g \omega_r$ at sensor location $\ell$
\begin{align}
\tau_g^{(\ell)}=\exp\left[-\sigma_g \rho_g^{(\ell)}/2\right],
\label{tau_g_dependence}
\end{align}
where the cross-section $\sigma_g$ may be gas dependent.

We consider the following two tasks: the estimation of the gradient and the frequency-dependent functions of gas densities.

\subsection{Estimating the density gradient of multiple gases}

We begin with gas-density-gradient estimation. For each species $g$, the parameter of interest is $\nabla_g\equiv\partial_x\rho_g(x)|_{x=x_0}$. We place $K$ sensors at $x_\ell=x_0+\ell\Delta x$, $\ell=1,\ldots,K$, with small uniform spacing $\Delta x$. The locally linear density profile gives
\begin{align}
\tau^{(\ell)}_g
&=\exp[-\sigma_g\rho_g(x_\ell)/2]
\simeq 1-\frac{\sigma_g}{2}
\left[\rho_g(x_0)+\ell\Delta x\,\nabla_g\right],
\label{eq:tau_ell}
\end{align}
where the approximation also uses weak absorption.

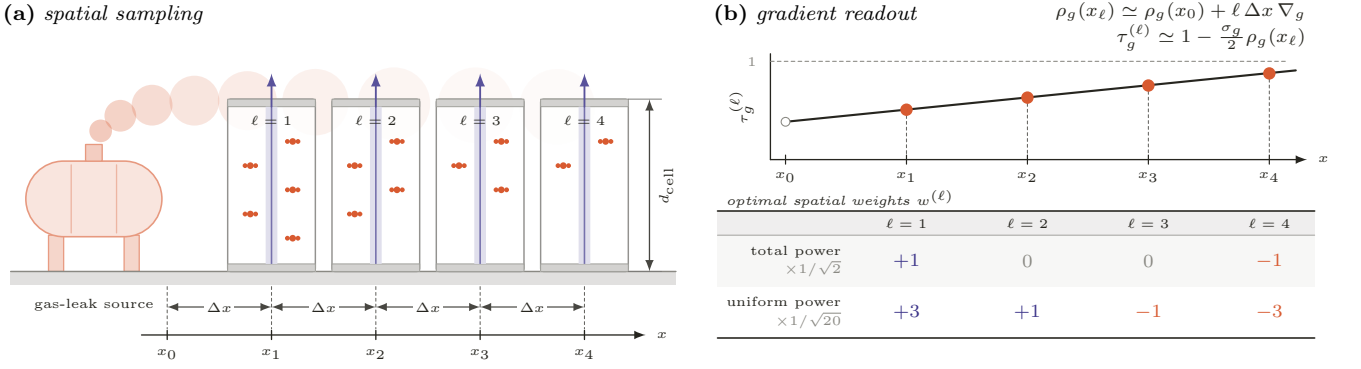
\begin{figure*}
    \centering
%
%
%
%

\definecolor{gastwo}{HTML}{D85A30}%
\definecolor{splitcol}{HTML}{3C3489}%
\definecolor{greyink}{HTML}{8A8983}%
\definecolor{darkink}{HTML}{26251F}%

\newcommand{\dgmol}[3]{%
  \begin{scope}[shift={(#1,#2)}]
    \draw[#3!75,line width=0.4pt] (-0.070,0)--(0.070,0);
    \fill[#3] (-0.070,0) circle (0.032);
    \fill[#3] (0.070,0) circle (0.032);
    \fill[#3] (0,0) circle (0.044);
  \end{scope}}%

\begin{tikzpicture}[x=1cm,y=1cm,line join=round,line cap=round]
\useasboundingbox (-0.10,0.02) rectangle (17.60,5.20);

\node[anchor=north west,inner sep=0pt,font=\footnotesize]
  at (-0.06,5.16) {\textbf{(a)} \textit{spatial sampling}};

\foreach \px/\py/\pr/\op in {1.24/3.48/0.15/0.32, 1.50/3.68/0.21/0.27,
                             1.92/3.82/0.27/0.21, 2.48/3.88/0.33/0.16,
                             3.18/3.88/0.39/0.12, 4.08/3.86/0.44/0.085,
                             5.12/3.84/0.48/0.060, 6.20/3.82/0.50/0.040,
                             7.28/3.80/0.50/0.028}
  \fill[gastwo,opacity=\op] (\px,\py) circle (\pr);

\fill[gastwo!25,draw=gastwo!60,line width=0.5pt] (0.55,1.62) rectangle (0.75,2.10);
\fill[gastwo!25,draw=gastwo!60,line width=0.5pt] (1.55,1.62) rectangle (1.75,2.10);
\fill[gastwo!16,draw=gastwo!62,line width=0.6pt,rounded corners=0.5cm]
  (0.25,2.08) rectangle (2.05,3.08);
\draw[gastwo!45,line width=0.5pt] (0.85,2.13)--(0.85,3.03);
\draw[gastwo!45,line width=0.5pt] (1.45,2.13)--(1.45,3.03);
\fill[gastwo!30,draw=gastwo!62,line width=0.5pt] (1.04,3.08) rectangle (1.26,3.30);
\node[font=\tiny,text=darkink] at (1.15,1.18) {gas-leak source};

\fill[greyink!26] (0.05,1.44) rectangle (8.85,1.62);
\draw[greyink,line width=0.55pt] (0.05,1.62)--(8.85,1.62);

\foreach \cx/\lbl/\nm in {3.50/1/5, 4.88/2/4, 6.26/3/3, 7.64/4/2}{
  \fill[white,opacity=0.80] (\cx-0.58,1.62) rectangle (\cx+0.58,3.90);
  \draw[darkink!50,line width=0.55pt,rounded corners=1.2pt]
    (\cx-0.58,1.62) rectangle (\cx+0.58,3.90);
  \fill[greyink!38,draw=greyink!80,line width=0.35pt]
    (\cx-0.58,1.62) rectangle (\cx+0.58,1.72);
  \fill[greyink!38,draw=greyink!80,line width=0.35pt]
    (\cx-0.58,3.80) rectangle (\cx+0.58,3.90);
  \draw[splitcol!16,line width=4.4pt,line cap=butt] (\cx,1.72)--(\cx,3.80);
  \draw[splitcol!70,line width=0.6pt,line cap=butt] (\cx,1.72)--(\cx,3.80);
  \draw[-latex,splitcol!85,line width=0.7pt] (\cx,3.80)--(\cx,4.24);
  \node[font=\tiny,text=darkink] at (\cx,3.62) {$\ell=\lbl$};
  \foreach \k in {1,...,\nm}{
    \pgfmathsetmacro{\my}{3.34-0.32*(\k-1)}
    \pgfmathsetmacro{\mx}{\cx+0.28*(2*mod(\k,2)-1)}
    \dgmol{\mx}{\my}{gastwo}}
}

\draw[darkink!60,line width=0.35pt] (8.22,1.62)--(8.66,1.62);
\draw[darkink!60,line width=0.35pt] (8.22,3.90)--(8.66,3.90);
\draw[latex-latex,darkink!85,line width=0.5pt] (8.50,1.62)--(8.50,3.90);
\node[font=\tiny,text=darkink,rotate=90] at (8.72,2.76) {$d_{\rm cell}$};

\foreach \gx in {2.12,3.50,4.88,6.26,7.64}
  \draw[darkink!70,dash pattern=on 1.1pt off 1.1pt,line width=0.4pt]
    (\gx,1.42)--(\gx,0.86);
\foreach \ax/\bx in {2.12/3.50,3.50/4.88,4.88/6.26,6.26/7.64}{
  \draw[latex-latex,darkink!85,line width=0.45pt] (\ax,1.15)--(\bx,1.15);
  \node[font=\tiny,text=darkink,fill=white,inner sep=0.8pt]
    at ({(\ax+\bx)/2},1.15) {$\Delta x$};}
\draw[-latex,darkink,line width=0.5pt] (1.78,0.78)--(8.45,0.78);
\node[font=\tiny,text=darkink,anchor=west] at (8.48,0.78) {$x$};
\foreach \tx/\tl in {2.12/0,3.50/1,4.88/2,6.26/3,7.64/4}{
  \draw[darkink,line width=0.5pt] (\tx,0.72)--(\tx,0.84);
  \node[font=\tiny,text=darkink] at (\tx,0.52) {$x_{\tl}$};}

\node[anchor=north west,inner sep=0pt,font=\footnotesize]
  at (9.34,5.16) {\textbf{(b)} \textit{gradient readout}};

\node[font=\scriptsize,text=darkink,anchor=east] at (17.30,5.06)
  {$\rho_g(x_\ell)\simeq\rho_g(x_0)+\ell\,\Delta x\,\nabla_g$};
\node[font=\scriptsize,text=darkink,anchor=east] at (17.30,4.72)
  {$\tau_g^{(\ell)}\simeq1-\tfrac{\sigma_g}{2}\rho_g(x_\ell)$};

\draw[darkink,line width=0.5pt] (10.10,3.10)--(10.10,4.50);
\draw[-latex,darkink,line width=0.5pt] (10.10,3.10)--(17.20,3.10);
\node[font=\tiny,text=darkink,anchor=west] at (17.23,3.10) {$x$};
\node[font=\tiny,text=darkink,rotate=90] at (9.72,3.80) {$\tau_g^{(\ell)}$};
\draw[greyink,dash pattern=on 1.4pt off 1.2pt,line width=0.4pt]
  (10.10,4.40)--(16.95,4.40);
\node[font=\tiny,text=greyink,anchor=east] at (10.03,4.40) {$1$};

\draw[darkink,line width=0.8pt] (10.30,3.60)--(17.05,4.28);
\foreach \mx/\my in {11.90/3.76, 13.50/3.92, 15.10/4.08, 16.70/4.24}{
  \draw[darkink!70,dash pattern=on 1.1pt off 1.1pt,line width=0.4pt]
    (\mx,\my)--(\mx,3.16);
  \fill[gastwo] (\mx,\my) circle (0.082);}
\draw[greyink,fill=white,line width=0.4pt] (10.30,3.60) circle (0.058);

\foreach \tx/\tl in {10.30/0,11.90/1,13.50/2,15.10/3,16.70/4}{
  \draw[darkink,line width=0.45pt] (\tx,3.04)--(\tx,3.16);
  \node[font=\tiny,text=darkink] at (\tx,2.87) {$x_{\tl}$};}

\node[font=\tiny\itshape,text=darkink,anchor=west] at (9.40,2.58)
  {optimal spatial weights $w^{(\ell)}$};

\fill[greyink!14] (9.40,2.10) rectangle (17.30,2.42);
\fill[greyink!07] (9.40,1.42) rectangle (17.30,2.10);
\draw[darkink,line width=0.6pt] (9.40,2.42)--(17.30,2.42);
\draw[greyink!75,line width=0.4pt] (9.40,2.10)--(17.30,2.10);
\draw[darkink,line width=0.6pt] (9.40,0.74)--(17.30,0.74);

\foreach \cx/\l in {11.90/1,13.50/2,15.10/3,16.70/4}
  \node[font=\tiny,text=darkink] at (\cx,2.26) {$\ell=\l$};

\node[font=\tiny,text=darkink,anchor=east,align=right] at (11.15,1.76)
  {total power\\[1pt] {\color{greyink}$\times1/\sqrt{2}$}};
\foreach \cx/\v/\cc in {11.90/{+1}/splitcol, 13.50/{0}/greyink,
                        15.10/{0}/greyink, 16.70/{-1}/gastwo}
  \node[font=\scriptsize,text=\cc] at (\cx,1.76) {$\v$};

\node[font=\tiny,text=darkink,anchor=east,align=right] at (11.15,1.08)
  {uniform power\\[1pt] {\color{greyink}$\times1/\sqrt{20}$}};
\foreach \cx/\v/\cc in {11.90/{+3}/splitcol, 13.50/{+1}/splitcol,
                        15.10/{-1}/gastwo, 16.70/{-3}/gastwo}
  \node[font=\scriptsize,text=\cc] at (\cx,1.08) {$\v$};

\end{tikzpicture}
    \caption{$K$-sample gas-gradient sensing, illustrated for $K=4$. (a) Identical local cells sample the gas field at $x_\ell=x_0+\ell\Delta x$ along $x$, while \QZ{each optical path has a fixed length} $d_{\rm cell}$ along $z\perp x$. Each distributed DCS node returns the local field transmissivity $\tau^{(\ell)}$. (b) Weak-gradient model and normalized optimal weights. The total-power optimum uses only the endpoints, whereas uniform per-sample power gives the least-squares slope. Blue and red indicate positive and negative estimator weights, respectively.}
    \label{fig:density_gradient_schematic}
\end{figure*}

Following the linear estimation setup, define the species-resolved functional
$\theta_{w,g}=\sum_{\ell=1}^K w^{(\ell)}\tau^{(\ell)}_g$, using the same spatial weight direction for every species. To eliminate the nuisance term $1-\sigma_g\rho_g(x_0)/2$, an admissible gradient weight must satisfy
$
\sum_{\ell=1}^{K}w^{(\ell)}=0 .
$
Then
\be
\theta_{w,g}\simeq
-\frac{\sigma_g\Delta x}{2}
\left(\sum_{\ell=1}^K\ell w^{(\ell)}\right)\nabla_g,
\ee
hence the unbiased estimator of $\nabla_g $ is
\be 
\hat \nabla_g=
-\frac{2}{\sigma_g\Delta x\sum_{\ell=1}^K\ell w^{(\ell)}}
\hat\theta_{w,g},
\qquad
\sum_{\ell=1}^K\ell w^{(\ell)}\neq0 .
\ee
Below we summarize the results of the estimation error optimization, while the details are in Appendix \ref{app:gradient_est}.

For the total power constrained case, \QZ{we use the optimized allocation in Eq.}~\eqref{eq:minvar_totalPower_opt}. \QZ{Define $\mathcal S_G=\sum_{n=-N}^{N}G_n^{-1}$. The optimized} total-power allocation gives
\be
{\rm var}\hat \nabla_g
\simeq \frac{16}{ A^2\sigma_g^2\Delta x^2}\mathcal S_G
\frac{\left(\sum_{\ell=1}^{K}|w^{(\ell)}|\right)^2}{\left(\sum_{\ell=1}^{K}\ell w^{(\ell)}\right)^2}.
\ee
where the total power constraint is $\hbar\Omega_c A^2/T=2P_m$, $\Omega_c$ is the carrier frequency, $T$ \QZ{is the integration time}.
This ratio is minimized by using only the two endpoints $w_{\rm tot}=(1,0,\cdots,0,-1)/\sqrt{2}$,
\be
{\rm var} \hat \nabla_g |_{\rm tot}
\simeq 
\frac{64}{A^2\sigma_g^2\Delta x^2(K-1)^2}\mathcal S_G
\label{eq:gradient_total_main}
\ee
Thus, under the total-power constraint, the optimal gradient readout sends sample-comb power only to the first and $K$th positions.

\QZ{For a uniform power constraint} over the $K$ samples, Eq.~\eqref{eq:minvar_uniform} gives
\be
{\rm var}\hat \nabla_g
\simeq \frac{16K}{A^2\sigma_g^2\Delta x^2}\mathcal S_G
\frac{\sum_{\ell=1}^{K}|w^{(\ell)}|^2}{\left(\sum_{\ell=1}^{K}\ell w^{(\ell)}\right)^2}.
\ee
where each sample is \QZ{subject to a power constraint} $\propto A^2/K$.
The unit-norm optimum is the least-squares slope direction
\be
w_{\rm uni}^{(\ell)}=\sqrt{\frac{3}{K(K^2-1)}}(K+1-2\ell),
\qquad \ell=1,\ldots,K,
\ee
which gives
\be
{\rm var}\hat \nabla_g|_{\rm uni}
\simeq 
\frac{192}{A^2\sigma_g^2\Delta x^2(K^2-1)}\mathcal S_G
\label{eq:gradient_uniform_main}
\ee
It is never lower than Eq.~\eqref{eq:gradient_total_main} for integer $K\ge 2$ as expected, \QZ{since the uniform power constraint is stricter}, while the difference is only \QZ{a constant factor of 3} at $K\to \infty$.

We emphasize that in this example, all species of gases share the identical optimal weight choice in distributing the quantum comb, \QZ{as the gradient can be estimated} from the same functional form. In the next section, we address a more complicated case where different species of gases require different functional forms.

\subsection{Estimating frequency-dependent functions}
\label{sec:multigas}

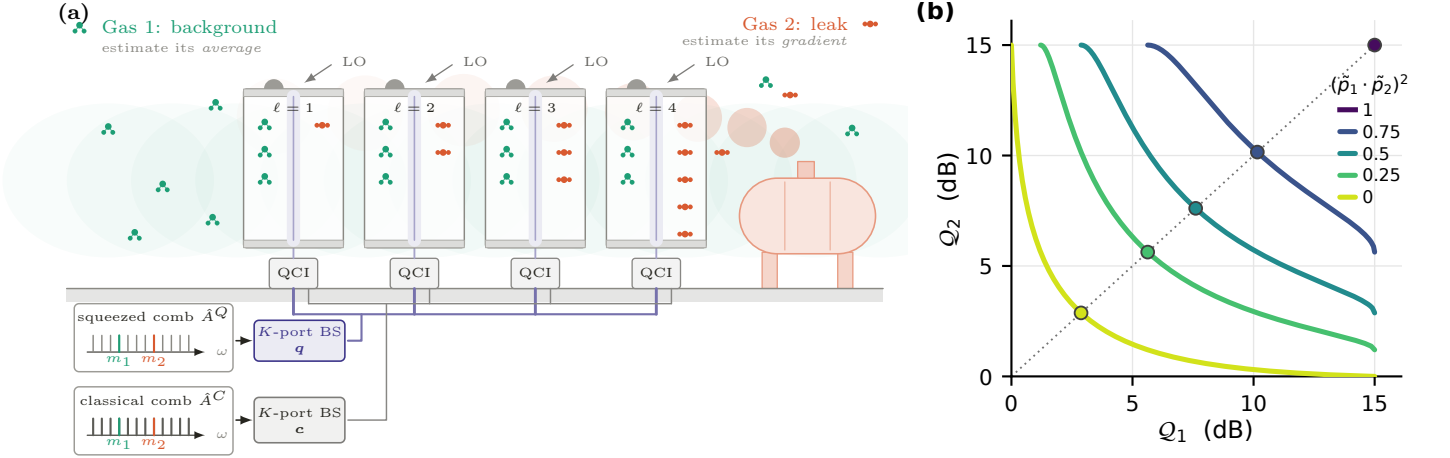
\begin{figure*}[t]
\centering
\begin{minipage}[c]{0.622\textwidth}\centering
%
%
%
%
%

\definecolor{gasone}{HTML}{1D9E75}
\definecolor{gastwo}{HTML}{D85A30}
\definecolor{splitcol}{HTML}{3C3489}
\definecolor{greyink}{HTML}{8A8983}
\definecolor{darkink}{HTML}{26251F}

\newcommand{\molbent}[3]{%
  \begin{scope}[shift={(#1,#2)}]
    \draw[#3!75,line width=0.4pt] (0,0.044)--(-0.060,-0.034)
                                  (0,0.044)--(0.060,-0.034);
    \fill[#3] (-0.060,-0.034) circle (0.032);
    \fill[#3] (0.060,-0.034) circle (0.032);
    \fill[#3] (0,0.044) circle (0.044);
  \end{scope}}

\newcommand{\mollin}[3]{%
  \begin{scope}[shift={(#1,#2)}]
    \draw[#3!75,line width=0.4pt] (-0.070,0)--(0.070,0);
    \fill[#3] (-0.070,0) circle (0.032);
    \fill[#3] (0.070,0) circle (0.032);
    \fill[#3] (0,0) circle (0.044);
  \end{scope}}

\begin{tikzpicture}[x=1cm,y=1cm,line join=round,line cap=round]
\useasboundingbox (-0.12,-0.68) rectangle (10.98,5.42);

\node[anchor=north west,inner sep=0pt,font=\bfseries\footnotesize]
  at (-0.08,5.40) {(a)};

\foreach \k in {0,...,12}{
  \pgfmathsetmacro{\hx}{0.60+0.85*\k}
  \fill[gasone,opacity=0.030] (\hx,3.05) ellipse (1.40 and 1.02);}
\foreach \mx/\my in {0.60/3.72, 1.32/2.96, 2.02/4.04, 0.95/2.32,
                     1.98/2.52, 9.30/4.34, 10.44/3.70}
  {\molbent{\mx}{\my}{gasone}}

\foreach \px/\py/\pr/\op in {9.55/3.55/0.20/0.30, 9.05/3.75/0.26/0.24,
                             8.40/3.88/0.32/0.18, 7.55/3.96/0.38/0.13,
                             6.50/4.00/0.44/0.09, 5.30/3.98/0.48/0.06,
                             4.00/3.94/0.50/0.04, 2.80/3.90/0.50/0.025}
  \fill[gastwo,opacity=\op] (\px,\py) circle (\pr);
\mollin{8.72}{3.42}{gastwo}
\mollin{9.62}{4.18}{gastwo}

\fill[gastwo!25,draw=gastwo!60,line width=0.5pt] (9.25,1.62) rectangle (9.45,2.10);
\fill[gastwo!25,draw=gastwo!60,line width=0.5pt] (10.25,1.62) rectangle (10.45,2.10);
\fill[gastwo!16,draw=gastwo!62,line width=0.6pt,rounded corners=0.5cm]
  (8.95,2.08) rectangle (10.75,3.08);
\draw[gastwo!45,line width=0.5pt] (9.55,2.13) -- (9.55,3.03);
\draw[gastwo!45,line width=0.5pt] (10.15,2.13) -- (10.15,3.03);
\fill[gastwo!30,draw=gastwo!62,line width=0.5pt] (9.72,3.08) rectangle (9.94,3.30);

\fill[greyink!22] (0.05,1.44) rectangle (10.85,1.62);
\draw[greyink,line width=0.55pt] (0.05,1.62) -- (10.85,1.62);

\foreach \cx/\lbl/\na/\nb in {3.05/1/3/1, 4.65/2/3/2, 6.25/3/3/3, 7.85/4/3/5}{
  \draw[rounded corners=1.2pt,darkink!55,line width=0.5pt,fill=greyink!10]
    (\cx-0.32,1.62) rectangle (\cx+0.32,2.02);
  \node[font=\tiny,text=darkink] at (\cx,1.82) {QCI};
  \draw[splitcol!45,line width=0.7pt] (\cx,2.02) -- (\cx,2.16);
  \fill[white,opacity=0.80] (\cx-0.66,2.16) rectangle (\cx+0.66,4.26);
  \draw[greyink,line width=0.55pt,rounded corners=1.2pt]
    (\cx-0.66,2.16) rectangle (\cx+0.66,4.26);
  \fill[greyink!30,draw=greyink!70,line width=0.35pt]
    (\cx-0.66,2.16) rectangle (\cx+0.66,2.26);
  \fill[greyink!30,draw=greyink!70,line width=0.35pt]
    (\cx-0.66,4.16) rectangle (\cx+0.66,4.26);
  \draw[splitcol!10,line width=5.0pt] (\cx,2.26) -- (\cx,4.16);
  \draw[splitcol!45,line width=0.6pt] (\cx,2.26) -- (\cx,4.16);
  \fill[greyink!85] (\cx-0.13,4.26) arc (0:180:0.13);
  \draw[-latex,darkink!60,line width=0.5pt] (\cx+0.50,4.58) -- (\cx+0.14,4.34);
  \node[font=\tiny,text=darkink!70,anchor=west] at (\cx+0.53,4.62) {LO};
  \node[font=\tiny,text=darkink] at (\cx,4.02) {$\ell=\lbl$};
  \foreach \k in {1,...,\na}{
    \pgfmathsetmacro{\my}{3.78-0.36*(\k-1)}
    \molbent{\cx-0.38}{\my}{gasone}}
  \foreach \k in {1,...,\nb}{
    \pgfmathsetmacro{\my}{3.78-0.36*(\k-1)}
    \mollin{\cx+0.38}{\my}{gastwo}}
}

\molbent{0.22}{5.06}{gasone}
\node[font=\scriptsize,text=gasone,anchor=west] at (0.40,5.06)
  {Gas 1: background};
\node[font=\tiny,text=greyink,anchor=west] at (0.40,4.79)
  {estimate its \emph{average}};
\mollin{10.68}{5.12}{gastwo}
\node[font=\scriptsize,text=gastwo,anchor=east] at (10.50,5.12)
  {Gas 2: leak};
\node[font=\tiny,text=greyink,anchor=east] at (10.50,4.88)
  {estimate its \emph{gradient}};

\draw[rounded corners=1.8pt,greyink!85,line width=0.55pt]
  (0.15,0.52) rectangle (2.25,1.42);
\node[font=\tiny,text=darkink] at (1.20,1.27) {squeezed comb $\hat A^{Q}$};
\draw[-latex,darkink,line width=0.4pt] (0.32,0.78) -- (1.88,0.78);
\foreach \k in {0,...,11}{
  \pgfmathsetmacro{\lx}{0.40+0.115*\k}
  \draw[greyink,line width=0.55pt] (\lx,0.78) -- (\lx,1.00);}
\draw[gasone,line width=0.8pt] (0.745,0.78) -- (0.745,1.00);
\draw[gastwo,line width=0.8pt] (1.205,0.78) -- (1.205,1.00);
\node[font=\tiny,text=gasone] at (0.745,0.66) {$m_1$};
\node[font=\tiny,text=gastwo] at (1.205,0.66) {$m_2$};
\node[font=\tiny,text=greyink,anchor=west] at (1.91,0.78) {$\omega$};

\draw[-latex,darkink,line width=0.6pt] (2.29,0.93) -- (2.51,0.93);
\draw[rounded corners=1.8pt,splitcol,line width=0.6pt,fill=splitcol!10]
  (2.53,0.66) rectangle (3.73,1.20);
\node[font=\tiny,text=splitcol,align=center] at (3.13,0.93)
  {$K$-port BS\\ $\bm q$};

\draw[splitcol!70,line width=0.8pt] (3.77,0.93) -- (3.95,0.93) -- (3.95,1.28);
\draw[splitcol!70,line width=0.8pt] (3.05,1.28) -- (7.85,1.28);
\foreach \cx in {3.05,4.65,6.25,7.85}
  \draw[splitcol!70,line width=1.1pt] (\cx,1.28) -- (\cx,1.62);

\draw[rounded corners=1.8pt,greyink!85,line width=0.55pt]
  (0.15,-0.58) rectangle (2.25,0.32);
\node[font=\tiny,text=darkink] at (1.20,0.17) {classical comb $\hat A^{C}$};
\draw[-latex,darkink,line width=0.4pt] (0.32,-0.32) -- (1.88,-0.32);
\foreach \k in {0,...,11}{
  \pgfmathsetmacro{\lx}{0.40+0.115*\k}
  \draw[darkink!75,line width=0.8pt] (\lx,-0.32) -- (\lx,-0.10);}
\draw[gasone,line width=0.8pt] (0.745,-0.32) -- (0.745,-0.10);
\draw[gastwo,line width=0.8pt] (1.205,-0.32) -- (1.205,-0.10);
\node[font=\tiny,text=gasone] at (0.745,-0.44) {$m_1$};
\node[font=\tiny,text=gastwo] at (1.205,-0.44) {$m_2$};
\node[font=\tiny,text=greyink,anchor=west] at (1.91,-0.32) {$\omega$};

\draw[-latex,darkink,line width=0.6pt] (2.29,-0.12) -- (2.51,-0.12);
\draw[rounded corners=1.8pt,darkink!60,line width=0.6pt,fill=greyink!12]
  (2.53,-0.42) rectangle (3.73,0.18);
\node[font=\tiny,text=darkink,align=center] at (3.13,-0.12)
  {$K$-port BS\\ $\bm c$};

\draw[darkink!55,line width=0.55pt] (3.77,-0.12) -- (4.28,-0.12) -- (4.28,1.42);
\draw[darkink!55,line width=0.55pt] (3.25,1.42) -- (8.05,1.42);
\foreach \cx in {3.25,4.85,6.45,8.05}
  \draw[darkink!55,line width=0.55pt] (\cx,1.42) -- (\cx,1.62);

\end{tikzpicture}%
\end{minipage}%
\hfill
\begin{minipage}[c]{0.368\textwidth}\centering
  \includegraphics{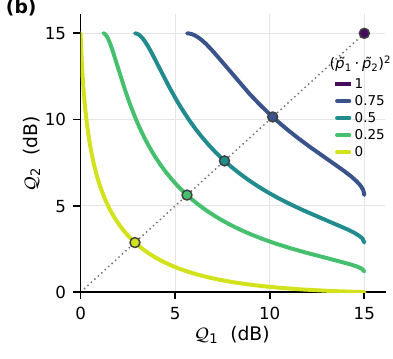}%
\end{minipage}
\caption{Distribution tradeoff for multiple gases.
(a)~One example of the setting: gas~1 is a diffuse background, whose average we take as a reference, while gas~2 escapes from a
leaking tank and we want its gradient in
order to locate the leak.  The same
treatment applies to any pair of linear functionals $\theta^{(g)}=\sum_\ell
w_g^{(\ell)}\rho_g^{(\ell)}$ of the density fields, and to any number $S$ of
species.
(b)~Equation~\eqref{eq:mg_frontier} at $G=15$~dB; reachable pairs lie below
and to the left of each curve, and dots mark $\mathcal Q_1=\mathcal Q_2=\mathcal Q_\star$.  Which curve applies is set by the overlap $\bigl(\tilde{\bm p}_1\cdot\tilde{\bm p}_2\bigr)^2$ of the two targets.  For the pair in
panel~(a), the two targets are orthogonal.}
\label{fig:mg}
\end{figure*}

Now we consider an example where we aim to estimate different functionals for different gases. As shown in Fig.~\ref{fig:mg}(a), this can correspond to a case where we want to estimate the average density of a background gas \QZ{`Gas 1' as a reference}, while aiming to estimate the gradient of another poisonous gas `Gas 2' to search for the leak. 

For each gas, we want to estimate its own functional
of the density field,
$\theta^{(g)}=\sum_\ell w_g^{(\ell)}\rho_g^{(\ell)}$. Eq.~\eqref{tau_g_dependence} \QZ{in the weak absorption limit}
gives $\tau_{m_g}^{(\ell)}-1\simeq-\tfrac12\sigma_g\rho_g^{(\ell)}$, so
$\hat\theta^{(g)}=-2\hat O_{m_g}/\sigma_g$ is unbiased with
${\rm var}\,\hat\theta^{(g)}=4{\rm Var}[\hat O_{m_g}]/\sigma_g^2$, with $\hat{O}_{m_g}$ given in Eq.~\eqref{eq:estimator}.
Distinct lines are statistically independent, since the noise of
$\hat O_m$ involves only the pairs
$(\hat\calA^{Q,(\ell)}_{n,m},\hat\calA^{Q,(\ell)}_{n,-m})$.

We consider the case where only frequency-independent beamsplitters are
available for distributing the quantum and classical combs.  The postprocessing
weights $v^{(\ell)}$ and the scalar comb envelope $A_{m}$ may be chosen per
line, but the per-sensor ratios are common to all lines,
$A_{m_g}^{(\ell)}=c^{(\ell)}A_{m_g}$.  Writing
$p_g^{(\ell)}=w_g^{(\ell)}/A_{m_g}^{(\ell)}$ for the effective weights of line
$m_g$, and $\tilde{\bm p}_g\equiv\bm p_g/\|\bm p_g\|$ for its direction,
Eq.~\eqref{eq:var_matrix_app} at $\tau_n\to1$ gives (Appendix~\ref{app:multigas})
\be
{\rm Var}[\hat O_{m_g}]
=4\bigl[M\|\bm p_g\|^2-(M-\mathcal S_G)(\bm p_g\cdot\bm q)^2\bigr],
\label{eq:mg_var}
\ee
with $M=2N+1$ the number of comb lines. Equation~\eqref{eq:mg_var} reduces to
Eq.~\eqref{eq:varO_qBSopt} at $\bm q=\tilde{\bm p}_g$, and its classical
benchmark is $\mathcal S_G\to M$ at the same $\{A_m^{(\ell)}\}$ and $\bm v$.
Hence, for uniform gain $G_n=G$, the quantum advantage for the functional of
gas $g$ is
\be
\mathcal Q_g\equiv
\frac{{\rm Var}^{\rm cl}[\hat O_{m_g}]}{{\rm Var}[\hat O_{m_g}]}
=\frac{1}{1-\bigl(1-1/G\bigr)\bigl(\tilde{\bm p}_g\cdot\bm q\bigr)^2} .
\label{eq:mg_adv}
\ee
Note that $\mathcal Q_g$ compares the same $\{A_m^{(\ell)}\}$ and $\bm v$ with
and without squeezing, so $\|\bm p_g\|^2$ cancels and no power budget enters
Eq.~\eqref{eq:mg_adv}.

The advantage is therefore controlled by a single direction cosine between the
beamsplitter weights $\bm q$ and the target direction $\tilde{\bm p}_g$, and is
maximal, $\mathcal Q_g=G$, only when the two are aligned.  A single
frequency-independent $\bm q$ must serve all $S$ lines, so when the targets
differ one cannot align it with every $\tilde{\bm p}_g$ at once, and the
advantages $\mathcal Q_1,\dots,\mathcal Q_S$ must be traded against one
another.  Since a unit $\bm q$ has a fixed total length to divide among the
targets, the achievable alignments $(\tilde{\bm p}_g\cdot\bm q)^2$ are confined
to an ellipsoid fixed by the target geometry [Eq.~\eqref{eq:mg_ellipsoid}].  For $S=2$ this
region is
\be
\bigl(\tilde{\bm p}_1\cdot\bm q\bigr)^2+\bigl(\tilde{\bm p}_2\cdot\bm q\bigr)^2
-2\bigl|\tilde{\bm p}_1\cdot\bm q\bigr|\,\bigl|\tilde{\bm p}_2\cdot\bm q\bigr|\,
\bigl|\tilde{\bm p}_1\cdot\tilde{\bm p}_2\bigr|
\le 1-\bigl(\tilde{\bm p}_1\cdot\tilde{\bm p}_2\bigr)^2 ,
\label{eq:mg_frontier}
\ee
and its boundary is the frontier plotted in
Fig.~\ref{fig:mg}(b).  Gaining advantage on one gas along this
boundary costs advantage on the other.  The symmetric choice
$\mathcal Q_1=\mathcal Q_2\equiv\mathcal Q_\star$ lies on the boundary at
$(\tilde{\bm p}_g\cdot\bm q)^2=(1+|\tilde{\bm p}_1\cdot\tilde{\bm p}_2|)/2$,
giving the best advantage that can be guaranteed to both gases,
\be
\mathcal Q_\star=\left[1-\bigl(1-1/G\bigr)
\frac{1+\bigl|\tilde{\bm p}_1\cdot\tilde{\bm p}_2\bigr|}{2}\right]^{-1} .
\label{eq:mg_sym}
\ee
Collinear targets give $\mathcal Q_\star=G$, the full advantage, whereas
orthogonal ones give $\mathcal Q_\star=2G/(G+1)$, which saturates at $3$~dB
however strong the squeezing.  Since
$\tilde{\bm p}_g\propto(w_g^{(\ell)}/c^{(\ell)})_\ell$ with $\bm c$ shared by
all species, a common $\bm c$ cannot make two targets collinear unless the
$\bm w_g$ already are, and the ceiling is a property of the sensing task.

We point out that the above trade-off between the quantum advantages for \QZ{different gases appears from} the frequency-independent beamsplitters. If a frequency-dependent beamsplitter is available, all frequency-separated species can enjoy maximal quantum advantage simultaneously. Frequency-dependent beamsplitters can potentially be engineered \QZ{with an inverse-design approach}~\cite{schul2026inverse}. 
 
\section{Discussions}

The proposed distributed dual-comb spectroscopy protocol applies to \QZ{both heterodyne and division receiver cases}. We have taken the weak absorption limit in the main text and \QZ{their performances are equivalent}; \QZ{in the finite-loss case, we expect} the same noise terms in the single sensor case~\cite{shi2025theory} will come into play, as we detail in the Supplementary Notes. The proposed distributed DCS scheme is compatible with both intra-comb-line squeezing (self-referred for division or cross-referred for heterodyne) and cross-comb-line entanglement~\cite{shi2025theory}, and the performance comparison between these choices follows from Ref~\cite{shi2025theory} directly, which we leave for future work.

\section{Acknowledgements}
This material is based upon work supported by the Air Force Office of Scientific Research under the PRISM MURI, award number FA9550-24-1-0349, and the Office of Naval Research under the DISCO-DEQS MURI, award number N000142612102. Any opinions, findings, and conclusions or recommendations expressed in this material are those of the author(s) and do not necessarily reflect the views of the United States Air Force nor the United States Navy.
 
\section{Methods}
\label{sec:endmatter_division}

\subsection{Distributed DCS with division receiver}
The heterodyne network measures absorption and phase together
and therefore inherits sensitivity to sample phase fluctuations. It also demands
cross-referenced squeezing, whose centre frequency is offset by
$n\Delta\omega_r$ from the bright comb lines. The division
receiver~\cite{herman2024squeezed,shi2025theory} trades the phase information for
\QZ{a readout that is self-calibrating against the unknown comb envelope, is insensitive}
to sample phase, and needs only self-referenced squeezing, centred on each comb's
own lines. Here we extend the QCI construction to this receiver.
 
\subsubsection{Architecture}
\label{sec:architecture_div}
 
\begin{figure}
    \centering
    \includegraphics[width=\linewidth]{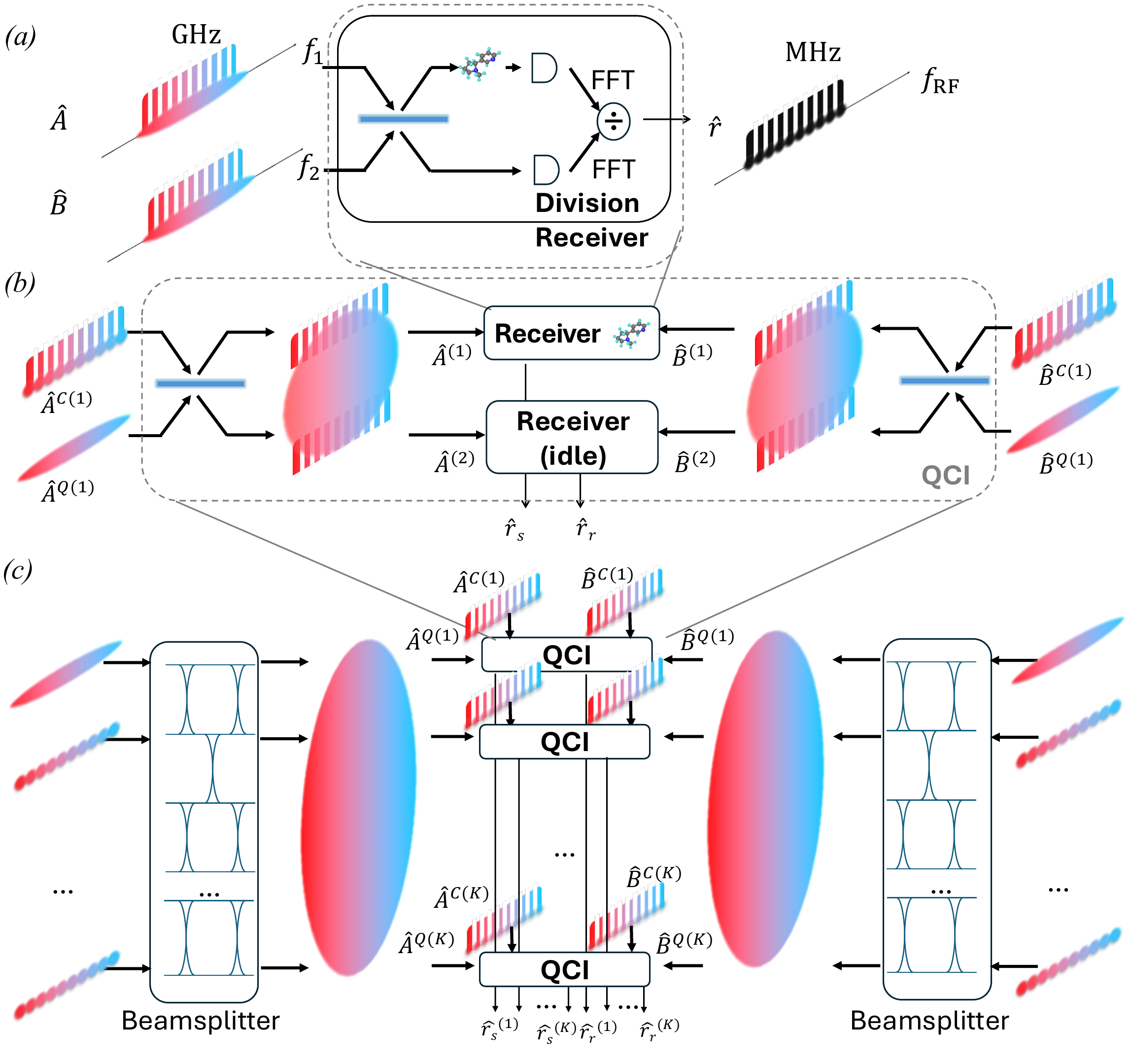}
    \caption{(a) Dual comb spectroscopy \QZ{with a division receiver; (b)} QCI yielding the ratios $\hat r_s$ and \QZ{$\hat r_r$; (c)} Distributed QCI for $K$ samples}
    \label{fig:division}
\end{figure}
 
The network (Fig.~\ref{fig:division}) is driven by four source combs: bright
classical combs $\hat A^{C},\hat B^{C}$ at repetition rates
$(\omega_r+\Delta\omega_r)/2\pi$ and $\omega_r/2\pi$ with per-line amplitudes
$A_n,B_n$, and zero-mean quantum combs $\hat A^{Q},\hat B^{Q}$ with self-referred intra comb-line squeezing:
\be
{\rm var}\bigl(\hat\calA^{Q}_{n,n+m}+\hat\calA^{Q\d}_{n,n-m}\bigr)=\frac{1}{G_{A,n}},
\quad
{\rm var}\bigl(\hat\calB^{Q}_{n,m}+\hat\calB^{Q\d}_{n,-m}\bigr)=\frac{1}{G_{B,n}} .
\label{eq:em_sources}
\ee
Each source is distributed to the $K$ sensors by a passive $K$-port array,
\bal
\hat\calA^{Q,(\ell)}_{n,m}&=q_A^{(\ell)}\hat\calA^{Q}_{n,m}
+\sum_{k\ge2}U_{Q_A}^{(\ell,k)}\hat e^{Q_A,(k)}_{n,m},
&A^{(\ell)}_n&=c^{(\ell)}A_n,
\nonumber\\
\hat\calB^{Q,(\ell)}_{n,m}&=q_B^{(\ell)}\hat\calB^{Q}_{n,m}
+\sum_{k\ge2}U_{Q_B}^{(\ell,k)}\hat e^{Q_B,(k)}_{n,m},
&B^{(\ell)}_n&=c^{(\ell)}B_n,
\label{eq:em_dist}
\eal
where $\bm q_A,\bm q_B,\bm c$ are the first columns of the array unitaries,
$\|\bm q_A\|=\|\bm q_B\|=\|\bm c\|=1$, and the $\hat e^{(k)}$ are the vacua
entering the unused ports.

At sensor $\ell$, two QCI beamsplitters form the branches
\be
\hat A^{(\ell)}_\pm=\frac{\hat A^{C,(\ell)}\pm\hat A^{Q,(\ell)}}{\sqrt2},
\qquad
\hat B^{(\ell)}_\pm=\frac{\hat B^{C,(\ell)}\pm\hat B^{Q,(\ell)}}{\sqrt2},
\label{eq:em_qci}
\ee
and each of the two branches is read by its own division receiver: a balanced
beamsplitter produces the arms
\be
\hat E^{\pm,(\ell)}_{1}=\frac{\hat A^{(\ell)}_\pm+\hat B^{(\ell)}_\pm}{\sqrt2},
\qquad
\hat E^{\pm,(\ell)}_{2}=\frac{\hat A^{(\ell)}_\pm-\hat B^{(\ell)}_\pm}{\sqrt2},
\label{eq:em_arms}
\ee
of which only $\hat E^{+,(\ell)}_{1}$ traverses the sample channel
$\calL^{(\ell)}$ of Eq.~\eqref{loss_channel}. All four arms are photodetected
with $\hat I^{+,(\ell)}_{j}(t)$ and $\hat I^{-,(\ell)}_{j}(t)$ as the photocurrent
operators of arm $j$ in the two branches, and the two branch ratios are
\be
\hat r^{(\ell)}_{\rm s}\equiv-\frac{\hat I^{+,(\ell)}_{1}(m\Delta\omega_r)}
{\hat I^{+,(\ell)}_{2}(m\Delta\omega_r)},
\qquad
\hat r^{(\ell)}_{\rm r}\equiv-\frac{\hat I^{-,(\ell)}_{1}(m\Delta\omega_r)}
{\hat I^{-,(\ell)}_{2}(m\Delta\omega_r)},
\label{eq:em_rdef}
\ee
for the sample branch and the reference branch respectively, with means
\be
r^{(\ell)}_{\rm s}=\kappa^{(\ell)}_m,
\qquad
r^{(\ell)}_{\rm r}=1,
\label{eq:em_ratio}
\ee
where $\kappa^{(\ell)}_n=|\tau^{(\ell)}_n|^2$, and we read out their difference
\be
\Delta\hat r^{(\ell)}\equiv\hat r^{(\ell)}_{\rm s}-\hat r^{(\ell)}_{\rm r} .
\label{eq:em_deltar}
\ee

We assume a real uniform envelope $A_n=A$, $B_n=B$ from here.

\subsubsection{Readout noise}
\label{sec:em_cancel}
 
Let $\hat P^{X}_n(s)\equiv\hat\calX_{n,s+m}+\hat\calX^{\d}_{n,s-m}$
and $\hat S_X(s)\equiv\sum_n\hat P^{X}_n(s)$.
\QZ{When each ratio in \eqref{eq:em_rdef} is linearised about its mean, each numerator
splits} into self-beat terms, in which a comb's fluctuations beat with its own
bright lines, and cross-beat terms, in which they beat with the other comb's
lines; the sample branch weights them by $\kappa_n+r_{\rm s}$ and
$\kappa_n-r_{\rm s}$ respectively, and we define
\be
g^{(\ell)}_n\equiv\frac{\kappa^{(\ell)}_n+r^{(\ell)}_{\rm s}}{2},
\qquad
h^{(\ell)}_n\equiv\frac{\kappa^{(\ell)}_n-r^{(\ell)}_{\rm s}}{2}.
\label{eq:em_gh}
\ee
 
The classical fluctuations enter the two branches with the same sign and the quantum fluctuations with opposite signs. The subtraction therefore acts oppositely on the two. Explicitly, suppressing the label
$(\ell)$ on $g_n,h_n$ and on the mode operators,
\begin{widetext}
    \bal
\Delta\bigl(\Delta\hat r^{(\ell)}\bigr)
=\frac{2}{c^{(\ell)}}\sum_n\Bigl\{
&\frac{1}{B}\Bigl[(g_n\!+\!1)\hat P^{A^Q}_n(n)+(g_n\!-\!1)\hat P^{A^C}_n(n)
+h_n\bigl(\hat P^{B^Q}_n(n)+\hat P^{B^C}_n(n)\bigr)\Bigr]
\nonumber\\
+\;&\frac{1}{A}\Bigl[(g_n\!+\!1)\hat P^{B^Q}_n(0)+(g_n\!-\!1)\hat P^{B^C}_n(0)
+h_n\bigl(\hat P^{A^Q}_n(0)+\hat P^{A^C}_n(0)\bigr)\Bigr]\Bigr\}
+\Delta\hat\calV ,
\label{eq:em_noise}
\eal
\end{widetext}
 
where $\Delta\hat\calV$ collects the vacuum admitted by the sample, weighted by
$\sqrt{\kappa_n(1-\kappa_n)}$ and local to each sensor.
 
Since $g_n\to1$ and $h_n\to0$ on transparent
lines and, more generally, for any uniform spectrum, the classical vacuum and the
cross-beats drop out there and
\be
\Delta\bigl(\Delta\hat r^{(\ell)}\bigr)\Big|_{\kappa\to1}
=\frac{4}{c^{(\ell)}}\left[\frac{\hat S^{(\ell)}_{A^Q}(n)}{B}
+\frac{\hat S^{(\ell)}_{B^Q}(0)}{A}\right] .
\label{eq:em_k1}
\ee

Away from this limit the $h_n$ terms read each quantum comb at the other centre,
where the mode pairing is mismatched \QZ{and its variance is amplified}, as we
quantify in Appendix~\ref{app:division}.
 
\subsubsection{Network estimator}
\label{sec:em_network}
 
Since $\expval{\Delta\hat r^{(\ell)}}=\kappa^{(\ell)}_m-1\simeq2(\tau^{(\ell)}_m-1)$,
$\hat O_m=\sum_\ell w^{(\ell)}\hat y^{(\ell)}$ is
unbiased for $\theta_m$ where $\hat y^{(\ell)}\equiv\Delta\hat r^{(\ell)}/2$. Its
fluctuation in the weak-absorption limit is
$\Delta\hat O_m=2\sum_\ell[p^{(\ell)}_B\hat S^{(\ell)}_{A^Q}(n)
+p^{(\ell)}_A\hat S^{(\ell)}_{B^Q}(0)]$, with the effective weights
\be
p^{(\ell)}_A=\frac{w^{(\ell)}}{c^{(\ell)}A},
\qquad
p^{(\ell)}_B=\frac{w^{(\ell)}}{c^{(\ell)}B}.
\label{eq:em_p}
\ee
 
Because the arrays act
identically on every frequency mode and their unused ports are vacuum,
\be
{\rm cov}\bigl(\hat S^{(\ell)}_X(s),\hat S^{(\ell')}_X(s)\bigr)
=M\delta_{\ell\ell'}+q_X^{(\ell)}q_X^{(\ell')}\bigl(V-M\bigr),
\label{eq:em_Scov}
\ee
with $M$ the number of comb lines and $V\equiv\sum_n\frac{1}{G_n}$. Hence
\be
{\rm var}\,\hat O_m^{\rm div}=F\bigl[\bm p_B,\bm q_A\bigr]+F\bigl[\bm p_A,\bm q_B\bigr],
\label{eq:em_var}
\ee
where $F[\bm p,\bm q]\equiv4\bigl[M\|\bm p\|^2-(M-V)(\bm p\cdot\bm q)^2\bigr]$, to
be compared with ${\rm var}\,\hat O^{\rm het}_m=F[\bm p_A,\bm q_A]$.
 
By Cauchy--Schwarz,
$(\bm p\cdot\bm q)^2\le\|\bm p\|^2$ at $\|\bm q\|=1$, so each term of
\eqref{eq:em_var} is minimised at
\be
\bm q_A=\frac{\bm p_B}{\|\bm p_B\|},
\quad
\bm q_B=\frac{\bm p_A}{\|\bm p_A\|},
\quad
{\rm var}\,\hat O^{\rm div}_m\to 4V\bigl(\|\bm p_A\|^2+\|\bm p_B\|^2\bigr) ,
\label{eq:em_qopt}
\ee
and by \eqref{eq:em_p} these two directions are parallel, so a single splitter
setting is simultaneously optimal. Next, $\|\bm p_X\|^2=X^{-2}\sum_\ell
|w^{(\ell)}|^2/|c^{(\ell)}|^2$ for $X=A,B$ is minimised over the classical array at
$|c^{(\ell)}|^2\propto|w^{(\ell)}|$, giving $\|\bm p_X\|^2=W/X^2$. Finally
$1/A^2+1/B^2$ is minimised under the exposure constraint
$A^2+B^2=4P_m$ at $A^2=B^2$, giving
$1/P_m$. Repeating the last two steps for the heterodyne network, whose
constraint is $\sum_\ell|A^{(\ell)}_m|^2=2P_m$,
\be
\min{\rm var}\,\hat O^{\rm div}_m=\frac{4VW}{P_m}
=2\,\min{\rm var}\,\hat O^{\rm het}_m ,
\quad
W\equiv\Bigl(\sum_\ell\bigl|w^{(\ell)}\bigr|\Bigr)^2
\label{eq:em_opt}
\ee
both attained at the same network configuration,
\be
\bigl|c^{(\ell)}\bigr|^2=\frac{\bigl|w^{(\ell)}\bigr|}{\sum_{\ell'}\bigl|w^{(\ell')}\bigr|},
\quad
q_A^{(\ell)}=q_B^{(\ell)}={\rm sgn}\bigl(w^{(\ell)}\bigr)\bigl|c^{(\ell)}\bigr|.
\label{eq:em_optpoint}
\ee

Setting $G_n=1$ gives $V\to M$ and the classical benchmark, so the advantage is
$\calQ=M/V\to G$ for uniform gain.
The heterodyne network needs one
squeezed source rather than two, its LO comb being classical, and that is what
buys it the factor of two in \eqref{eq:em_opt}.

\section{Reducing the detector number}

\begin{figure}
    \centering
    \includegraphics[width=\linewidth]{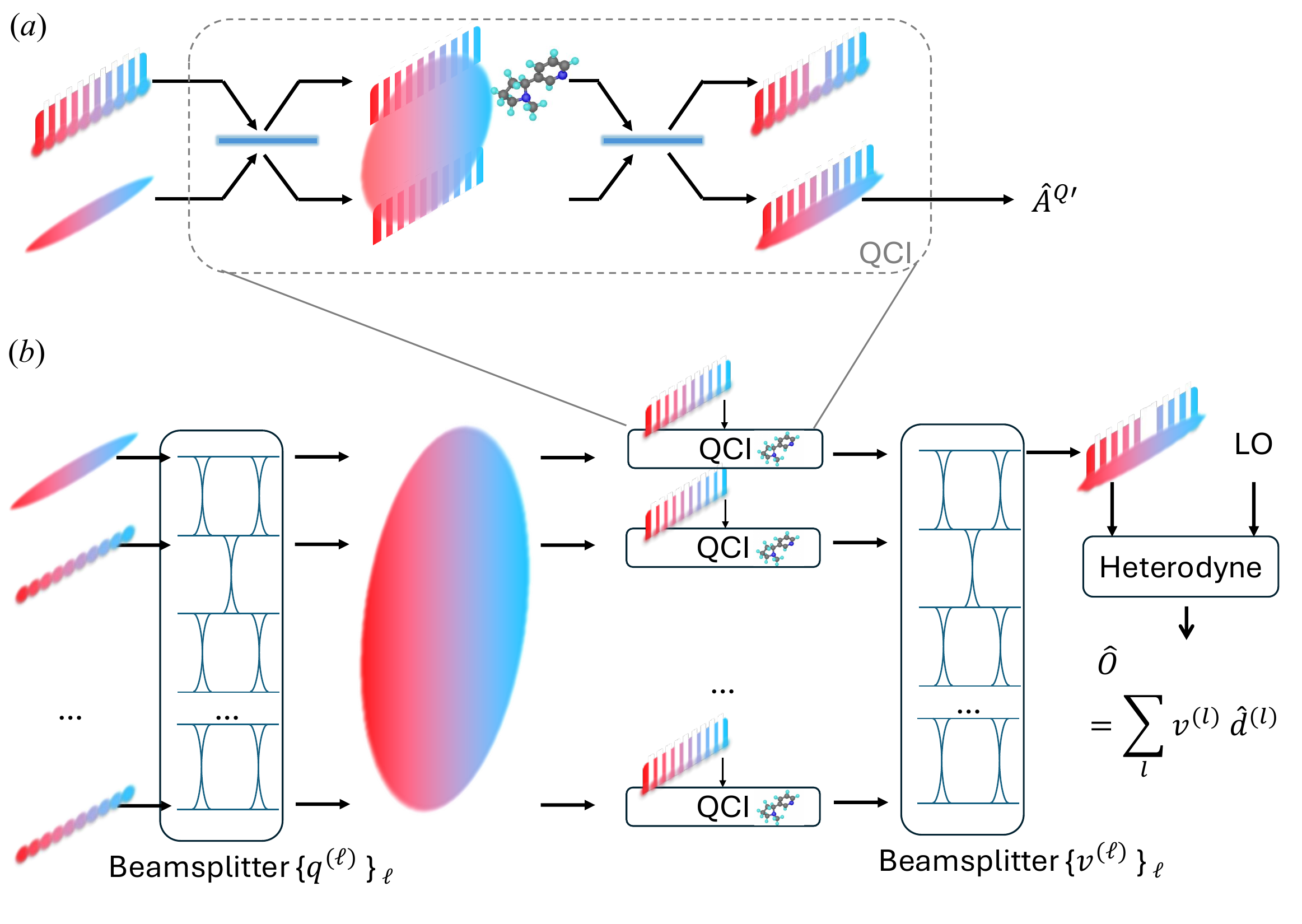}
    \caption{(a) Quantum comb interferometry (QCI) \QZ{based on a Mach-Zehnder interferometer}, without heterodyne detection; (b) Distributed QCI for $K$ samples, with the $K$ outputs combined by a beamsplitter before the heterodyne detection.}   
    \label{fig:schematic_concept_BScombined}
\end{figure}

The schematic in Fig.~\ref{fig:schematic_concept} can be implemented alternatively using only one heterodyne detector, \QZ{via combining the distributed signals} before heterodyne, as shown in Fig.~\ref{fig:schematic_concept_BScombined}. The heterodyne detector design is identical to Fig.~\ref{fig:schematic_concept}(a), while \QZ{we postpone the heterodyne in QCI Fig.~\ref{fig:schematic_concept}(b) until after} a beamsplitter recombination as shown in Fig.~\ref{fig:schematic_concept_BScombined}(c). Such coherent recombination concentrates the informative photons into one output mode and reduces the number of detectors to one.

%

\appendix 
\onecolumngrid

\section{Full analysis with arbitrary absorption}

\label{app:MZI_onlyMinusOutput}

\subsection{Summary of the results}

\begin{figure}[htbp]
    \centering
    \includegraphics[width=.8\linewidth]{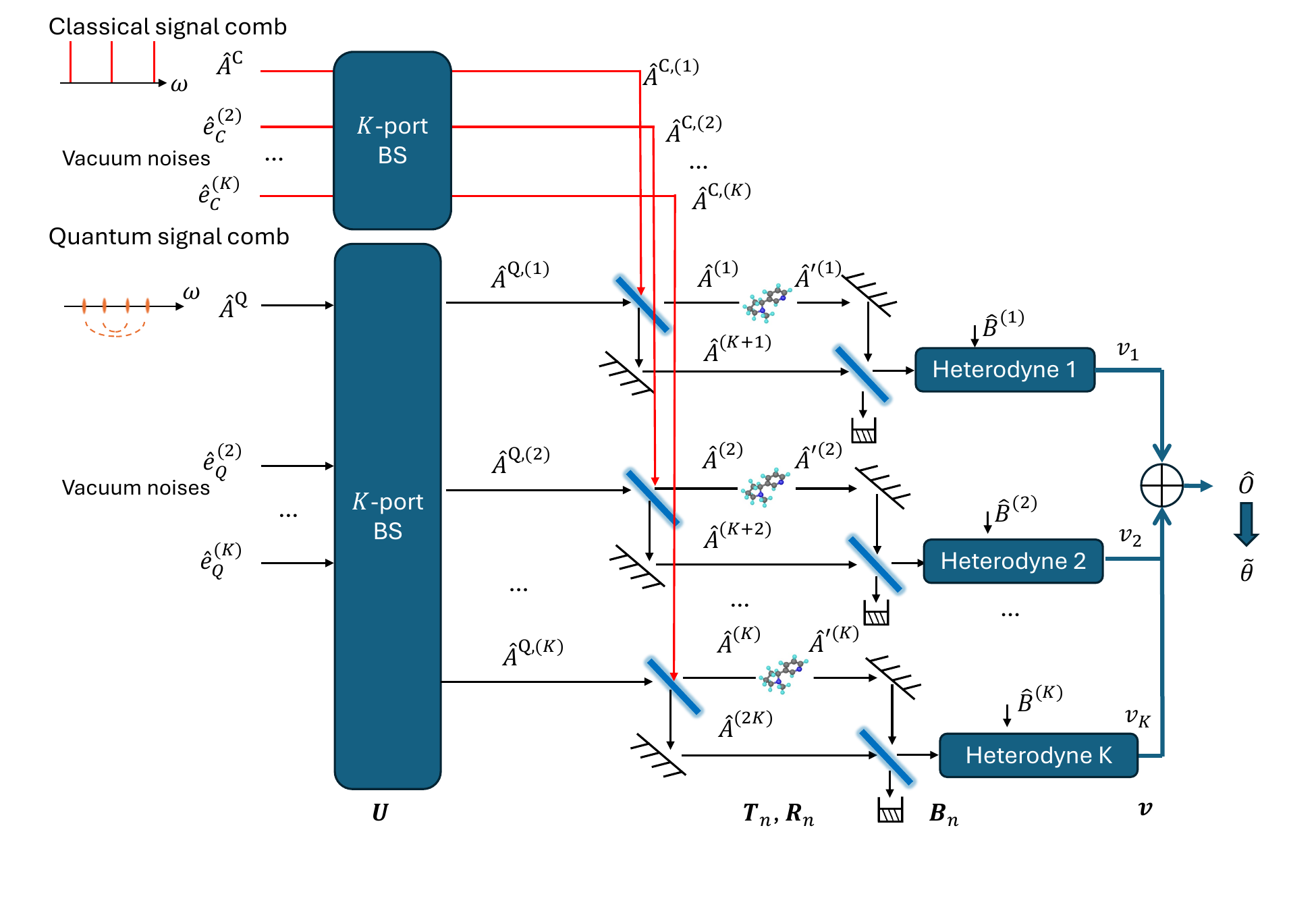}
    \caption{Schematic of the $K$-sample distributed dual-comb spectroscopy based on Mach-Zehnder interferometry (MZI) with reduced outputs. Only the $K$ destructive MZI outputs of the total $2K$ outputs are measured, which recovers the quantum probes and \QZ{extracts the differential signal} from classical probes. The total power is constrained by the power of $\hat A^C$.}
    \label{fig:schematic_MZI_reduced}
\end{figure}

Consider the $K$-sample $2N+1$-line distributed dual-comb heterodyne interferometry as shown in Fig.~\ref{fig:schematic_MZI_reduced}. The estimator variance formula is
\be 
{\rm var} \hat \theta = \sum_{n=-N}^N \sigma^2 (\{\tau_n^{(j)}\}_{j=1}^K)  
\ee
where we define the variance contribution from the $n$th line as
\bal 
\sigma^2(\{\tau_n^{(j)}\}_{j=1}^K) &=
|\sum_{j=1}^{K} \frac{1}{2} v^{(j)}   B_n^{(j)}  (\tau_n^{(j)}+1)   q^{(j)} |^2 \cdot \frac{1}{G_n} 
+   \sum_{k=2}^{K} |\sum_{j=1}^{K} \frac{1}{2}  v^{(j)}  B_n^{(j)}   (\tau_n^{(j)}+1)  U_Q^{(j,k)}|^2 \\
&\qquad \qquad +  \sum_{k=1}^{K} |\sum_{j=1}^{K} \frac{1}{2}  v^{(j)}  B_n^{(j)}   (\tau_n^{(j)}-1)  U_C^{(j,k)}|^2 
+ \frac{1}{2}|\sum_{j=1}^{K} v^{(j)}    B_n^{(j)}   \sqrt{1-|\tau_n^{(j)}|^2}    |^2 \,.
\eal  
Here $U_Q,U_C$ are the unitary matrices representing the distribution beamsplitters of the quantum and classical combs respectively. Specifically, \QZ{their first columns correspond} to the distribution of the squeezed quantum comb $\hat \calA^Q$ and the classical power $P_m$, as $U_Q^{(j,1)}= q^{(j)}$, $U_C^{(j,1)}=|A_m^{(\ell)}|/\sqrt{P_m}$. The factors $\tau_n^{(j)}\pm 1$ should be kept as complex quantities if sample phases are not compensated, whereas the loss noise depends only on $|\tau_n^{(j)}|^2$.
Full derivations of the formula are in Appendix \ref{app:SNR} as Eq.~\eqref{eq:var_app}.

Consider few-line absorption, where only $\eta\cdot 2 N\ll 2N$ lines \QZ{within the line set} $\mathcal I$ have absorption, i.e. $\{\kappa_n^{(j)}\}_{n\in \mathcal I}$, $|\mathcal I|=2\eta N$,  $\eta\ll 1$. Then 
\bal 
{\rm var} \hat \theta = (1-\eta)2N  \sigma^2 (\{1\}_{j=1}^K) + O(\eta)
\eal 
where the lossless-regime variance $\sigma^2(\{1\}_{j=1}^K)$ is given by Eq.~\eqref{eq:varO_unbiased}, which is optimized to \QZ{Eqs.~\eqref{eq:varO_qBSopt} and \eqref{eq:varO_opt}}.

\subsection{Full model}
\label{app:full_model}

\QZ{Here we consider the reduced} MZI protocol shown in Fig.~\ref{fig:schematic_MZI_reduced}, while in Appendix~\ref{app:MZI_fulloutput} we will formulate the full MZI model with all $2K$ outputs accessible. It is based on the interferences between the two probe resources: Classical signal combs with excited comb lines of high mean values but vacuum-limited fluctuations, and zero-mean quantum combs with squeezing or entanglement engineered, in the same spirit of dark port squeezing in LIGO~\cite{aasi2013enhanced}. For $K$-sample interferometry under the constraint on total power (dominated by the classical signal combs), we model the $K$ classical signal combs as distributed from a single input holding all the power via a $K$-port beamsplitter.

\QZ{Consider an $M$-line comb}, with $N$ pairs around \QZ{a center line at the carrier frequency}, $M=2N+1$. First we prepare a single classical comb 
\bal 
\hat A^{C}(t)&=\frac{1}{\sqrt{T}}\sum_{n=-N}^N\sum_{m=-2N}^{2N} e^{-i (n\omega_r+m\Delta\omega_r) t} (A_n\delta_{n,m} + \hat \calA_{n,m}^{{\rm in},C})\,,
\label{eq:classical_input}
\eal
which has a constrained power $P\propto |A|^2$.
Then we distribute it into $K$ classical comb probes by a $K$-port beamsplitter which fulfills the symplectic transform matrix $\bm U_C:=[\bm c,\bm u_C^{(2)}, \ldots ,\bm u_C^{(n)}]$:
\bal 
\hat A^{C,(j)}(t)= c^{(j)} \hat A^C(t) +  \sum_{k=2}^K U_C^{(j,k)} \hat e_C^{(k)}(t)
\label{eq:classical_probe}
\eal
where $\hat e_C^{(n)}$ are independent environment fields in vacuum. Here, we use the superscripts to index the input/output ports of the beamsplitter, and $(j,k)$ for matrix elements to denote the transfer from the $k$th input to the $j$th output in the spatial domain, to be discriminated from the frequency labels $n,m$ in the subscripts denoting frequency $n\omega_r+m\Delta \omega_r$. 

Similarly, the distributed quantum probe is generated from a single quantum comb 
\bal 
\hat A^{Q}(t)&=\frac{1}{\sqrt{T}}\sum_{n=-N}^N\sum_{m=-2N}^{2N} e^{-i (n\omega_r+m\Delta\omega_r) t}  \hat \calA_{n,m}^{{\rm in},Q},
\label{eq:quantum_input}
\eal 
where $\hat A^Q $ is in the squeezed state such that ${\rm var} ( \hat A_{n,-m}^{{\rm in},Q\dagger} +  \hat A_{n,m}^{{\rm in},Q })=1/G_n$ with squeezing gain $G_n$.
And it is distributed by another $K$-port BS with the transfer matrix $\bm U_Q=[\bm q,\bm u_Q^{(2)}, \ldots ,\bm u_Q^{(n)}]$
\bal 
\hat A^{Q,(j)}= q^{(j)} \hat A^Q +  \sum_{n=2}^K U_Q^{(j,n)} \hat e_Q^{(n)}
\label{eq:quantum_probe}
\eal
where $\hat e_Q^{(n)}$ are independent environment modes in vacuum states.

Meanwhile, we prepare the $K$ LO combs 
\be 
\hat B^{(j)}(t)=\frac{1}{\sqrt{T}}\sum_{n=-N}^N  e^{-i n\omega_r t}  B^{(j)}_{n}
\label{eq:LO comb}
\ee 
for the later use in heterodyne measurements. Here $B^{(j)}_{n}$ is the LO amplitude at the \QZ{$n$th comb line. We assume} that the LO is sufficiently strong $B^{(j)}_{n}\gg A_n$ so that the \QZ{quantum noise in the LO} is negligible.

\QZ{The first beamsplitter of the MZI} fulfills
\bal 
\hat A^{(k)}&= \frac{1}{\sqrt 2} ( \hat A^{C,(k)} + \hat A^{Q,(k)}  )\\
\hat A^{(k+K)}&= \frac{1}{\sqrt 2} ( \hat A^{C,(k)} - \hat A^{Q,(k)} )
\eal
for $1\le k\le K$. 
The transform from the $2K$ initial inputs $\hat {\bm A^{C, \rm in }}\equiv [\hat A^{C}, \hat v^{C,(2)}, \ldots, \hat v^{C,(K)}], \hat {\bm A^{Q, \rm in }}\equiv [\hat A^{Q}, \hat v^{Q(2)}, \ldots, \hat v^{Q(K)}]$ to the $2K$ MZI beamsplitter outputs $\hat A^{(1)}, \ldots , \hat A^{(2K)}$ is 
\bal 
\hat { A}^{(j)} = \frac{1}{\sqrt{2}}  U^{(j,k)}_C    \hat  A^{C, {\rm in} (k)} + \frac{1}{\sqrt{2}} U^{(j,k)}_Q    \hat  A^{Q, {\rm in} (k)}\\
\eal
for $1\le j\le K$,
\be 
\hat { A}^{(j)} = \frac{1}{\sqrt{2}}  U^{(j,k)}_C    \hat  A^{C, {\rm in} (k)} - \frac{1}{\sqrt{2}} U^{(j,k)}_Q    \hat  A^{Q, {\rm in} (k)}
\ee
for $K+1\le j\le 2K$. It can be compactly described by the $K\times K$ unitary matrices $\bm U_C$, $\bm U_Q$ as below
\be 
\hat {\bm A} = 
\bp 
\frac{1}{\sqrt{2}}
\left(\bm U_C \hat {\bm A}^{C,{\rm in}} + \bm U_Q \hat {\bm A}^{Q,{\rm in}} \right)
\\
\frac{1}{\sqrt{2}}
\left(\bm U_C \hat {\bm A}^{C,{\rm in}} - \bm U_Q \hat {\bm A}^{Q,{\rm in}} \right)
\ep 
\ee
We are specifically interested in the first column vectors 
$\bm c, \bm q$
of $\bm U_C, \bm U_Q$ as defined in \QZ{Eqs.~\eqref{eq:classical_probe} and \eqref{eq:quantum_probe}}, because $\bm c, \bm q$ are the distribution ratios of the classical signal comb $\hat A^C$ and quantum signal comb $\hat A^Q$ towards the MZI.  

In the middle of the MZI, the samples invoke loss, which implements \QZ{an array of bosonic loss channels in the frequency domain} with a field transmissivity spectrum $\tau_n$ at the neighborhood of frequency $n\omega_r$, with power transmissivity $\kappa_n=|\tau_n|^2$. We denote the frequency component of $\hat A^{\prime (j)}$ at frequency $n\omega_r+m\Delta \omega_r$ as $\hat A_{n,m}^{\prime (j)}$. Then
\bal 
\hat A_{n,m}^{\prime (j)}&= \tau_n^{(j)}  \hat A_{n,m}^{(j)}  + \sqrt{1-|\tau_n^{(j)}|^2} \hat e_{n,m}^{(j)}
\eal
for $1\le j\le K$, where the idler arms are lossless for $K+1\le j\le 2K$. The transform can be compactly described by 
\bal 
\hat {\bm A}'_{n,m}&=\bm T_n \hat {\bm A}_{n,m} +  \bm R_n \hat {\bm e}_{n,m} \\
&= 
\bp 
\frac{1}{\sqrt{2}}
\bm T_n \left(\bm U_C \hat {\bm A}^{C,{\rm in}} + \bm U_Q \hat {\bm A}^{Q,{\rm in}} \right)
\\
\frac{1}{\sqrt{2}}\left(\bm U_C \hat {\bm A}^{C,{\rm in}} - \bm U_Q \hat {\bm A}^{Q,{\rm in}} \right)
\ep 
+
\bp 
\bm R_n \hat e_{n,m}^{(j)}\\
0
\ep 
\eal
where the transmissivity matrix $\bm T_n = {\rm diag} \{\tau_n^{(1)}, \ldots ,\tau_n^{(K)}\}$ and reflectivity (loss) matrix $\bm R_n = {\rm diag} \{\sqrt{1-|\tau_n^{(1)}|^2}, \ldots ,\sqrt{1-|\tau_n^{(K)}|^2}\}$ are diagonal.

Finally, we apply the second beamsplitter of a standard MZI and discard the first $K$ \QZ{outputs which form constructive interference} that recovers the classical probe and destroys the quantum probe. We keep the last $K$ destructive interference \QZ{outputs which recover the quantum probe and keep} the differential signal from the classical probe. After the beamsplitter, the $K$ differential outputs are
\bal 
\hat {\bm A}_{n,m}^{\rm out}=&
\frac{1}{2}
\left[ (\bm T_n -\bm I) \bm U_C \hat {\bm A}_{n,m}^{C,{\rm in}} + (\bm T_n + \bm I) \bm U_Q \hat {\bm A}_{n,m}^{Q,{\rm in}} \right]\\
&+
\frac{1}{\sqrt{2} } 
\bm R_n \hat e_{n,m}^{(j)}
\eal
where $\bm I$ \QZ{is the $K\times K$ identity matrix}.
After heterodyne measurement, the final observable is 
\be 
\hat O_m=\sum_{j=1}^{K} v^{(j)} I^{(j)}_m
\ee 
at RF frequency $m\Delta \omega_r$, where 
\be 
I^{(j)}_m=\sum_{n=-N}^N (B_n^{(j)} \hat A_{n,-m}^{{\rm out}(j)\dagger} + B_n^{(j)*} \hat A_{n,m}^{{\rm out}(j)})
\ee 
is the photocurrent readout of the $j$th heterodyne measurement with the $j$th heterodyne LO amplitude being $B_n^{(j)} $. 

Overall, the input-output relation can be compactly described by
\be 
\hat {\bm O}_m=\bm v \cdot \sum_{n=-N}^N (\bm B_n \hat {\bm A}_{n,-m}^{{\rm out}\dagger} + \bm B_n^{*} \hat {\bm A}_{n,m}^{{\rm out}})
\ee
\QZ{where the postprocessing weight is $\bm v=[v^{(1)}, \ldots,v^{(2K)} ]^T$, and the diagonal matrix} $\bm B_n={\rm diag}\{B_n^{(1)}, \ldots, B_n^{(K)}\}$ describes the LO amplitudes.
\QZ{Consider a real signal} $A_n=A_n^{*}$ for simplicity and thus $B_n^{(j)}=B_n^{(j)*}$:
\be 
\hat {\bm O}_m=\bm v \cdot  \sum_{n=-N}^N \bm B_n ( \hat {\bm A}_{n,-m}^{{\rm out}\dagger} + \hat {\bm A}_{n,m}^{{\rm out}})
\ee

Below we consider the \QZ{estimation of the unknown parameter} $\theta=\sum_{j=1}^{K} w^{(j)} \tau_m^{(j)}$ for a specific $m$.

\subsection{SNR analyses}
\label{app:SNR}






\QZ{For the total-power-fixed case}, the constraint is on the classical comb amplitude $\sum_{n=-N}^N |A_{n}|^2= PT/\hbar \Omega_C$, where $\Omega_C$ is the carrier frequency that is much larger than the comb bandwidth.

Note that 
\be 
\expval {\hat O_m}= \sum_{j=1}^{K} v^{(j)}  B_n^{(j)}   \tau_m^{(j)}   c^{(j)}  \frac{A_m}{2} -  \sum_{j=1}^{K} v^{(j)}  B_n^{(j)}    c^{(j)}  \frac{A_m}{2}
\ee
To get an unbiased estimator $\expval{\hat \theta}=\theta$, we can choose 
\be 
\hat \theta=\hat O_m +  \sum_{j=1}^{K} v^{(j)}  B_n^{(j)}    c^{(j)}  \frac{A_m}{2}
\ee
with 
\be 
v^{(j)} B_n^{(j)} c^{(j)} \frac{A_m}{2}=w^{(j)}
\label{eq:varmin_cond_v}
\ee
for $1\le j\le K$. 

In this case, the estimator variance is
\bal 
{\rm var} \hat \theta &=
\sum_{n=-N}^N \Bigg( |\sum_{j=1}^{K} \frac{1}{2} v^{(j)}   B_n^{(j)}  (\tau_n^{(j)}+1)   q^{(j)} |^2 \cdot \frac{1}{G_n} 
+   \sum_{k=2}^{K} |\sum_{j=1}^{K} \frac{1}{2}  v^{(j)}  B_n^{(j)}   (\tau_n^{(j)}+1)  U_Q^{(j,k)}|^2 \\
&\qquad \qquad +  \sum_{k=1}^{K} |\sum_{j=1}^{K} \frac{1}{2}  v^{(j)}  B_n^{(j)}   (\tau_n^{(j)}-1)  U_C^{(j,k)}|^2 
+ \frac{1}{2}|\sum_{j=1}^{K} v^{(j)}    B_n^{(j)}   \sqrt{1-|\tau_n^{(j)}|^2}    |^2 \Bigg)\\
\label{eq:var_app}
\eal 
It can be written in a compact matrix product form 
\bal 
{\rm var} \hat \theta &=
\sum_{n=-N}^N \left(  \frac{1}{4} | \bm v\cdot  \bm   B_n ({\bm T}_{n}+\bm I)  \cdot    {\bm q} |^2 \cdot \frac{1}{G_n} 
+  \sum_{k=2}^{K}  \frac{1}{4} |\bm v\cdot  \bm   B_n ({\bm T}_{n}+\bm I)  \cdot   \bm u_Q^{(k)}|^2 
+ \sum_{k=1}^{K}  \frac{1}{4} |\bm v\cdot  \bm   B_n ({\bm T}_{n}-\bm I)  \cdot   \bm u_C^{(k)}|^2
+ \frac{1}{2}  | \bm v\cdot  \bm   B_n \bm R_{n}   |^2 \right)
\label{eq:var_matrix_app}
\eal 
Without loss of generality, we assume $B_n^{(j)}=1$ since it can be absorbed into $v^{(j)}$.
To allow frequency dependence we define $\bm v_n: v_n^{(j)}=v^{(j)}B_n^{(j)}$. 
This is necessary because one \QZ{cannot make $\bm q$ frequency-dependent}.

\subsubsection{Total power constraint}

\QZ{For the total-power-fixed case}, the constraint is on the classical comb amplitude $ |A_{m}|^2= 2P_mT/\hbar \Omega_C$, where $\Omega_C$ is the carrier frequency that is much larger than \QZ{the comb bandwidth. The total power} $P_m$ is then distributed towards the sensors, and we optimize the distribution ratios $\bm c$ here.

\QZ{The estimator variance is lower bounded by the Cauchy-Schwarz inequality, and the minimum is achieved at} $ q^{(j)}= \frac{v_n^{(j)}   (1+\tau_n^{(j)})   }{\sqrt{\sum_{j=1}^{K} | v_n^{(j)} (1+\tau_n^{(j)})  |^2}}$ for $1\le j\le K$, and due to the orthogonality of $\bm U$, $  \sum_{j=1}^{K}v_n^{(j)} (1+\tau_n^{(j)})   U_Q^{(j,k)}=0$ for any $k\neq 1$. 
Thus given $\bm v_n$ chosen by Eq.~\eqref{eq:varmin_cond_v}, we have $ q^{(j)}=\frac{w^{(j)}(1+\tau_n^{(j)}) / c^{(j)}  }{\sqrt{\sum_{j=1}^{K} | w^{(j)} (1+\tau_n^{(j)}) / c^{(j)}   |^2}}$ for $1\le j\le K$. Here we note that $ \bm v_n $ must be tuned to eliminate the frequency-$n$ dependence of $ v_n^{(j)} (1+\tau_n^{(j)} )$, because $\bm q$ is frequency independent.

\QZ{In the neighborhood of the sample lossless limit} $\tau^{(j)}\to 1$, we have
\be 
\min {\rm var} \hat \theta =  \sum_{n=-N}^N  \sum_{j=1}^{K} |v^{(j)}|^2    =  \sum_{n=-N}^N \frac{4}{A_m^2 G_n}  \sum_{j=1}^{K} |\frac{w^{(j)}}{ c^{(j)}}|^2   
\label{eq:minvar_totalPower}
\ee
We can further optimize $ c^{(j)}$ by setting $c^{(j)}=\sqrt{|w^{(j)}|}/\sqrt{\sum_{j=1}^{K} |w^{(j)}|}$ \QZ{according to the solution of the Lagrangian multiplier method; then} 
\be 
\min {\rm var} \hat \theta = \sum_{n=-N}^N  \frac{4}{A_m^2 G_n}   \left( \sum_{j=1}^{K} |w^{(j)}|\right)^2
\label{eq:minvar_totalPower_opt}
\ee

\subsubsection{Uniform power constraint}
Plugging in a uniform power allocation vector $c^{(j)}=1/\sqrt{K}$ in Eq.~\eqref{eq:minvar_totalPower}, we have 
\be 
\min {\rm var} \hat \theta = \sum_{n=-N}^N \frac{4K}{A_m^2 G_n}  \sum_{j=1}^{K} |w^{(j)}|^2\,.   
\label{eq:minvar_uniform}
\ee

\subsection{$K$-sample gradient estimation at weak-absorption limit}
\label{app:gradient_est}

Equations~\eqref{eq:minvar_totalPower_opt} and~\eqref{eq:minvar_uniform} estimate a linear functional of the amplitude transmissivities,
$\theta_{w,g}=\sum_{j=1}^{K}w^{(j)}\tau_g^{(j)}$. Consider $K\ge2$ sensors at $x_j=x_0+j\Delta x$:
\bal
\rho_g(x_j)&\simeq\rho_g(x_0)+j\Delta x\,\nabla_g,\\
\tau_g^{(j)}
&=\exp[-\sigma_g\rho_g(x_j)/2]
\simeq 1-\frac{\sigma_g}{2}
\left[\rho_g(x_0)+j\Delta x\,\nabla_g\right],
\qquad j=1,\ldots,K,
\eal
where $\nabla_g\equiv\partial_x\rho_g(x)|_{x=x_0}$. The weight vector must satisfy
\be
\sum_{j=1}^{K}w^{(j)}=0
\ee
to remove the position-independent terms. For the absorption-only gradient considered here we take the phase of $\tau_g^{(j)}$ to be zero or already compensated. Thus,
\be
\theta_{w,g}\simeq
-\frac{\sigma_g\Delta x}{2}
\left(\sum_{j=1}^{K}j w^{(j)}\right)\nabla_g.
\ee
An unbiased local estimator of the gradient is therefore
\be
\hat \nabla_g=
-\frac{2}{\sigma_g\Delta x\sum_{j=1}^{K}j w^{(j)}}
\hat \theta_{w,g},
\qquad
\sum_{j=1}^{K}j w^{(j)}\neq0 .
\ee
Defining $\mathcal S_G=\sum_{n=-N}^{N}G_n^{-1}$, Eq.~\eqref{eq:minvar_totalPower_opt} gives
\be
{\rm var}\hat \nabla_g(w)
\simeq \frac{16}{A^2\sigma_g^2\Delta x^2}\mathcal S_G
\frac{\left(\sum_{j=1}^{K}|w^{(j)}|\right)^2}
{\left(\sum_{j=1}^{K}j w^{(j)}\right)^2}.
\ee
The ratio is invariant under rescaling of $\bm w$. Using $\sum_{j=1}^{K}w^{(j)}=0$,
\be
\left|\sum_{j=1}^{K}j w^{(j)}\right|
=\left|\sum_{j=1}^{K}\left(j-\frac{K+1}{2}\right)w^{(j)}\right|
\le\frac{K-1}{2}\sum_{j=1}^{K}|w^{(j)}|,
\ee
and therefore
\be
\frac{\left(\sum_{j=1}^{K}|w^{(j)}|\right)^2}
{\left(\sum_{j=1}^{K}j w^{(j)}\right)^2}
\ge\frac{4}{(K-1)^2}.
\ee
Equality requires the two signs to be supported at the opposite endpoints. The unit-norm optimum and its variance are
\be
w_{\rm tot}^{(j)}=\frac{\delta_{j1}-\delta_{jK}}{\sqrt{2}},
\qquad
{\rm var}\hat \nabla_g^{\rm tot}
\simeq \frac{64}{A^2\sigma_g^2\Delta x^2(K-1)^2}\mathcal S_G .
\ee

For uniform power over all $K$ samples, Eq.~\eqref{eq:minvar_uniform} gives
\be
{\rm var}\hat \nabla_g(w)
\simeq \frac{16K}{A^2\sigma_g^2\Delta x^2}\mathcal S_G
\frac{\sum_{j=1}^{K}|w^{(j)}|^2}
{\left(\sum_{j=1}^{K}j w^{(j)}\right)^2}.
\ee
The zero-sum constraint and the Cauchy--Schwarz inequality give
\be
\left(\sum_{j=1}^{K}j w^{(j)}\right)^2
\le
\left[\sum_{j=1}^{K}\left(j-\frac{K+1}{2}\right)^2\right]
\sum_{j=1}^{K}|w^{(j)}|^2,
\qquad
\sum_{j=1}^{K}\left(j-\frac{K+1}{2}\right)^2
=\frac{K(K^2-1)}{12}.
\ee
Equality holds for the least-squares slope direction. The unit-norm optimum and its variance are
\be
w_{\rm uni}^{(j)}=\sqrt{\frac{3}{K(K^2-1)}}(K+1-2j),
\qquad
{\rm var}\hat \nabla_g^{\rm uni}
\simeq \frac{192}{A^2\sigma_g^2\Delta x^2(K^2-1)}\mathcal S_G .
\ee

\section{Multi-species distribution trade-off}
\label{app:multigas}

We derive Eq.~\eqref{eq:mg_var} and give the general-$S$ form of the trade-off
summarised in Sec.~\ref{sec:multigas}.

\subsection{Readout variance at weak absorption}
\label{app:mg_var}

At $\tau_n^{(j)}\to1$ one has $\bm T_n\to\bm I$, so
$\bm T_n+\bm I\to2\bm I$, $\bm T_n-\bm I\to0$ and $\bm R_n\to0$, and
Eq.~\eqref{eq:var_matrix_app} keeps only the quantum-comb terms,
\be
{\rm var}\,\hat\theta
=\sum_{n=-N}^{N}\left[\frac{|\bm v_n\cdot\bm q|^2}{G_n}
+\sum_{k=2}^{K}\bigl|\bm v_n\cdot\bm u_Q^{(k)}\bigr|^2\right].
\ee
Since $\{\bm q,\bm u_Q^{(2)},\ldots,\bm u_Q^{(K)}\}$ is an orthonormal basis,
$\sum_{k\ge2}|\bm v_n\cdot\bm u_Q^{(k)}|^2=\|\bm v_n\|^2-|\bm v_n\cdot\bm q|^2$,
so
\be
{\rm var}\,\hat\theta
=\sum_{n=-N}^{N}\left[\|\bm v_n\|^2
-\Bigl(1-\frac{1}{G_n}\Bigr)\bigl(\bm v_n\cdot\bm q\bigr)^2\right].
\label{eq:mg_var_general}
\ee
The unbiasedness condition $v^{(\ell)}A_m^{(\ell)}=2w^{(\ell)}$ gives
$\bm v=2\bm p_g$ on line $m_g$, frequency independent, and summing the $M$
lines with $\mathcal S_G=\sum_n G_n^{-1}$ yields Eq.~\eqref{eq:mg_var}.  No
power constraint has been used: Eq.~\eqref{eq:mg_var} holds for arbitrary
$\{A_m^{(\ell)}\}$, the constraints of Appendix~\ref{app:SNR} entering only
through the choice of $\bm c$.  Substituting the total-power optimum
$c^{(\ell)}=\sqrt{|w^{(\ell)}|}/\sqrt{\sum_{\ell'}|w^{(\ell')}|}$ gives
$\|\bm p_g\|^2=(\sum_\ell|w^{(\ell)}|)^2/A_{m_g}^2$ and recovers
Eq.~\eqref{eq:minvar_totalPower_opt} at $\bm q=\tilde{\bm p}_g$, while the
uniform allocation $c^{(\ell)}=1/\sqrt K$ gives
$\|\bm p_g\|^2=K\sum_\ell|w^{(\ell)}|^2/A_{m_g}^2$ and recovers
Eq.~\eqref{eq:minvar_uniform}.  Both return $\mathcal Q_g=G$ at
$\bm q=\tilde{\bm p}_g$, since $\|\bm p_g\|^2$ cancels from the ratio
Eq.~\eqref{eq:mg_adv}.

\subsection{Attainable advantages for $S$ species}
\label{app:mg_frontier}

Write $\mu_g\equiv(\tilde{\bm p}_g\cdot\bm q)^2\in[0,1]$ for the alignments, so
that Eq.~\eqref{eq:mg_adv} reads
$\mathcal Q_g=[1-(1-1/G)\mu_g]^{-1}$ and $\mathcal Q_g$ increases with $\mu_g$.
Let $\bm P=[\tilde{\bm p}_1,\dots,\tilde{\bm p}_S]$ and let
$\bm\Omega=\bm P^{T}\bm P$ be the Gram matrix of the target directions, with
$\bm\Omega^{+}$ its pseudoinverse, equal to $\bm\Omega^{-1}$ when the
$\tilde{\bm p}_g$ are linearly independent.  Since $\|\bm q\|=1$, the question
is which vectors of projections $\bm z=\bm P^{T}\bm q$, whose squares
$\mu_g=z_g^2$ are the alignments, a unit $\bm q$ can produce.  Any $\bm q$
realising a given $\bm z$ \QZ{costs at least as much as the least-norm solution}
$\bm q=\bm P\bm\Omega^{+}\bm z$, whose squared norm is
$\bm z^{T}\bm\Omega^{+}\bm z$, which therefore confines the alignments to an
ellipsoid set by the target geometry,
\be
\bm z^{T}\bm\Omega^{+}\bm z\le1 .
\label{eq:mg_ellipsoid}
\ee
Its boundary is attained and is the Pareto frontier of
$(\mathcal Q_1,\dots,\mathcal Q_S)$.

For $S=2$, writing $\Lambda\equiv(\tilde{\bm p}_1\cdot\tilde{\bm p}_2)^2$,
Eq.~\eqref{eq:mg_ellipsoid} reduces to
\be
\mu_1+\mu_2-2\sqrt{\mu_1\mu_2\Lambda}\le1-\Lambda ,
\label{eq:mg_frontier_app}
\ee
which is Eq.~\eqref{eq:mg_frontier} of the main text.  Setting
$\mu_1=\mu_2=\mu$ gives $2\mu(1-\sqrt\Lambda)=(1-\sqrt\Lambda)(1+\sqrt\Lambda)$,
hence the balanced point $\mu=(1+\sqrt\Lambda)/2$ used in
Eq.~\eqref{eq:mg_sym}.  For orthogonal targets Eq.~\eqref{eq:mg_ellipsoid}
becomes, at any $S$, the alignment sum rule $\sum_g\mu_g\le1$, i.e.
\be
\sum_{g=1}^{S}\frac{1}{\mathcal Q_g}\ \ge\ S-1+\frac1G ,
\label{eq:mg_sumrule}
\ee
with equality on the frontier.

More generally, the best advantage guaranteed to every species,
$\mathcal Q_\star=\max_{\bm q}\min_g\mathcal Q_g$, follows from
\bal
\frac1\Xi&\equiv\max_{\|\bm q\|=1}\ \min_g\ \mu_g
=\Bigl[\min\bigl\{\|\bm q\|^2:(\tilde{\bm p}_g\cdot\bm q)^2\ge1\ \forall
g\bigr\}\Bigr]^{-1},\\
\mathcal Q_\star&=\frac{\Xi}{\Xi-1+1/G}
\ \xrightarrow{\ G\to\infty\ }\ \frac{\Xi}{\Xi-1} .
\label{eq:mg_xi}
\eal
Here $\Xi\ge1$ measures how incompatible the targets are, with equality iff the
$\tilde{\bm p}_g$ are collinear, when a single $m$-independent configuration
returns $\mathcal Q_g=G$ for all $g$; otherwise $\mathcal Q_\star$ saturates at
$\Xi/(\Xi-1)$ however strong the squeezing.  For $S=2$,
$\Xi=2/(1+\sqrt\Lambda)$, reproducing Eq.~\eqref{eq:mg_sym}.  For targets with
pairwise disjoint supports the overlaps vanish termwise for every $\bm c$, and
Eq.~\eqref{eq:mg_sumrule} applies.

\section{Full analysis with both MZI outputs}
\label{app:MZI_fulloutput}

\begin{figure}
    \centering
    \includegraphics[width=.8\linewidth]{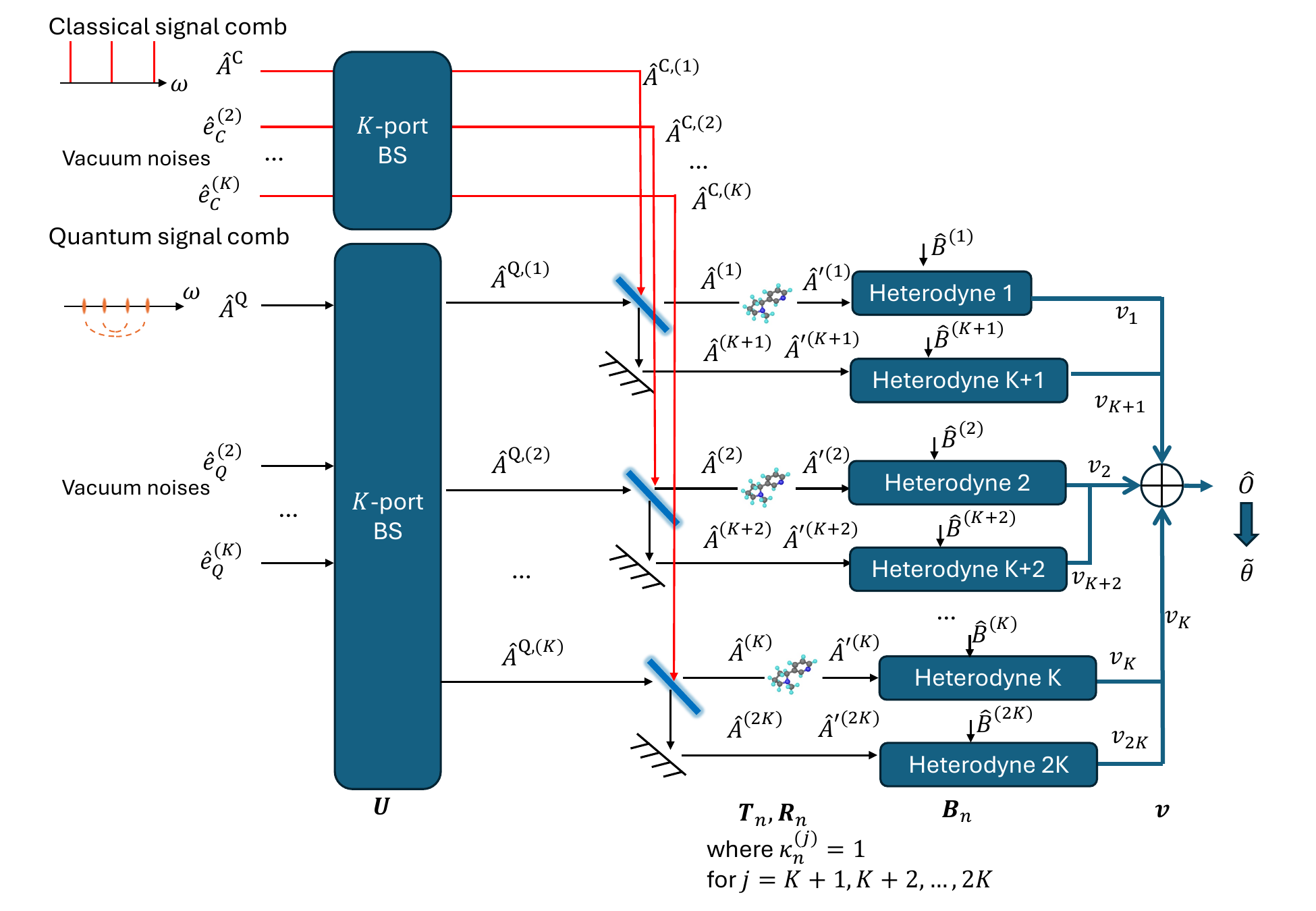}
    \caption{Schematic of the $K$-sample distributed dual-comb spectroscopy based on Mach-Zehnder interferometry \QZ{under a total power constraint}, with the full $2K$ outputs measured. }
    \label{fig:schematic_MZI_full}
\end{figure}

In Appendix \ref{app:MZI_onlyMinusOutput} we considered the MZI with only the destructive interference output collected, while the constructive interference output is discarded. Here we consider the MZI with both of the two outputs collected, as shown in Fig.~\ref{fig:schematic_MZI_full}.

\QZ{Consider an $M$-line comb}, with $N$ pairs around \QZ{a center line at the carrier frequency}, $M=2N+1$. The distributed classical probe is generated from a single classical comb 
\bal 
\hat A^{C}(t)&=\frac{1}{\sqrt{T}}\sum_{n=-N}^N\sum_{m=-2N}^{2N} e^{-i (n\omega_r+m\Delta\omega_r) t} (A_n\delta_{n,m} + \hat \calA_{n,m}^{{\rm in},C})
\label{eq:classical_input_app}
\eal 
which has a constrained power $P\propto |A|^2$.
The classical comb is distributed by a $K$-port BS which fulfills the symplectic transform matrix $\bm U_C=[\bm c,\bm u_C^{(2)}, \ldots ,\bm u_C^{(n)}]$:
\bal 
\hat A^{C,(j)}= c^{(j)} \hat A^C +  \sum_{n=2}^K U_C^{(j,n)} \hat e_C^{(n)}
\label{eq:classical_probe_app}
\eal
where $\hat e_C^{(n)}$ are independent environment modes in vacuum states. Here we use the superscripts $(j,k)$ to denote the transfer from the $k$th input to the $j$th output in the spatial domain, to be discriminated from the frequency labels $n,m$ in the subscripts denoting frequency $n\omega_r+m\Delta \omega_r$. For simplicity, we will omit the subscripts $n,m$ when the formula is frequency-independent.

Similarly, the distributed quantum probe is generated from a single quantum comb 
\bal 
\hat A^{Q}(t)&=\frac{1}{\sqrt{T}}\sum_{n=-N}^N\sum_{m=-2N}^{2N} e^{-i (n\omega_r+m\Delta\omega_r) t}  \hat \calA_{n,m}^{{\rm in},Q}
\label{eq:quantum_input_app}
\eal 
$\hat A^Q $ is in the squeezed state such that ${\rm var} ( \hat A_{n,-m}^{{\rm in},Q\dagger} +  \hat A_{n,m}^{{\rm in},Q })=1/G_n$ with squeezing gain $G_n$.
And it is distributed by another $K$-port BS with the transfer matrix $\bm U_Q=[\bm q,\bm u_Q^{(2)}, \ldots ,\bm u_Q^{(n)}]$
\bal 
\hat A^{Q,(j)}= q^{(j)} \hat A^Q +  \sum_{n=2}^K U_Q^{(j,n)} \hat e_Q^{(n)}
\label{eq:quantum_probe_app}
\eal
where $\hat e_Q^{(n)}$ are independent environment modes in vacuum states.

\QZ{The first beamsplitter of the MZI} fulfills
\bal 
\hat A^{(k)}&= \frac{1}{\sqrt 2} ( \hat A^{C,(k)} + \hat A^{Q,(k)}  )\\
\hat A^{(k+K)}&= \frac{1}{\sqrt 2} ( \hat A^{C,(k)} - \hat A^{Q,(k)} )
\eal
for $1\le k\le K$. 
The transform from the $2K$ initial inputs $\hat {\bm A^{C, \rm in }}\equiv [\hat A^{C}, \hat v^{C,(2)}, \ldots, \hat v^{C,(K)}], \hat {\bm A^{Q, \rm in }}\equiv [\hat A^{Q}, \hat v^{Q(2)}, \ldots, \hat v^{Q(K)}]$ to the $2K$ MZI beamsplitter outputs $\hat A^{(1)}, \ldots , \hat A^{(2K)}$ is 
\bal 
\hat { A}^{(j)} = \frac{1}{\sqrt{2}}  U^{(j,k)}_C    \hat  A^{C, {\rm in} (k)} + \frac{1}{\sqrt{2}} U^{(j,k)}_Q    \hat  A^{Q, {\rm in} (k)}\\
\eal
for $1\le j\le K$,
\be 
\hat { A}^{(j)} = \frac{1}{\sqrt{2}}  U^{(j,k)}_C    \hat  A^{C, {\rm in} (k)} - \frac{1}{\sqrt{2}} U^{(j,k)}_Q    \hat  A^{Q, {\rm in} (k)}
\ee
for $K+1\le j\le 2K$. 
It can be compactly described by the $K\times K$ unitary matrices $\bm U_C$, $\bm U_Q$ in the form of a block matrix as below
\be 
\hat {\bm A} = 
\frac{1}{\sqrt{2}}
\bm U
\bp \hat {\bm A}^{C,{\rm in}} \\ \hat {\bm A}^{Q,{\rm in}}\ep
\ee
where 
\be 
\bm U=\bp 
\bm U_C & \bm U_Q\\
\bm U_C &-\bm U_Q
\ep \,.
\ee
We are specifically interested in the first column vectors 
$\bm c, \bm q$
of $\bm U_C, \bm U_Q$ as defined in \QZ{Eqs.~\eqref{eq:classical_probe_app} and \eqref{eq:quantum_probe_app}}, because $\bm c, \bm q$ are the distribution ratios of the classical signal comb $\hat A^C$ and quantum signal comb $\hat A^Q$ towards the MZI.  For later convenience, we define the distribution ratios of $\hat A^C$ and $\hat A^Q$ towards the sample as
\be 
\tilde {\bm c}=
\bp 
\bm c/\sqrt{2} \\
\bm c/\sqrt{2}
\ep
\ee
\be 
\tilde {\bm q}=
\bp 
\bm q/\sqrt{2} \\
-\bm q/\sqrt{2}
\ep
\ee
 (note the minus sign in $\tilde {\bm q}$ \QZ{due to the BS phase}).

In the middle of the MZI, the samples invoke loss, which implements \QZ{an array of bosonic loss channels in the frequency domain} with a field transmissivity spectrum $\tau_n$ at the neighborhood of frequency $n\omega_r$, with $\kappa_n=|\tau_n|^2$. We denote the frequency component of $\hat A^{\prime (j)}$ at frequency $n\omega_r+m\Delta \omega_r$ as $\hat A_{n,m}^{\prime (j)}$. Then
\bal 
\hat A_{n,m}^{\prime (j)}&= \tau_n^{(j)}  \hat A_{n,m}^{(j)}  + \sqrt{1-|\tau_n^{(j)}|^2} \hat e_{n,m}^{(j)}
\eal
for $1\le j\le 2K$, where the idler arms are lossless $\tau^{(j)}=1$ for $K+1\le j\le 2K$. 

Finally, we skip the second beamsplitter of a standard MZI and directly make the heterodyne measurement, since the beamsplitter commutes with the measurement and can be fulfilled \QZ{by postprocessing with a weighted sum} with weights $v_1,\ldots, v_{2K}$. The final observable is 
\be 
\hat O_m=\sum_{j=1}^{2K} v^{(j)} I^{(j)}_m
\ee 
at RF frequency $m\Delta \omega_r$, where 
\be 
I^{(j)}_m=\sum_{n=-N}^N (B_n^{(j)} \hat A_{n,-m}^{\prime(j)\dagger} + B_n^{(j)*} \hat A_{n,m}^{\prime(j)})
\ee 
is the photocurrent readout of the $j$th heterodyne measurement with the $j$th heterodyne LO amplitude being $B_n^{(j)} $. 

Overall
\bal 
\hat O_m&=\sum_{j=1}^{2K} v^{(j)} \sum_{n=-N}^N (B_n^{(j)} \hat A_{n,-m}^{\prime(j)\dagger} + B_n^{(j)*} \hat A_{n,m}^{\prime(j)})\\
\eal 
Consider a phase-compensated real signal $A_n=A_n^{*}$ for simplicity and thus $B_n^{(j)}=B_n^{(j)*}$:
\bal
\hat O_m&=\sum_{j=1}^{2K} v^{(j)}    \tau^{(j)} \sum_{n=-N}^N  B_n^{(j)}  ( \hat A_{n,-m}^{(j)\dagger} + \hat A_{n,m}^{(j)}) + {\rm vac.}\\
\eal
where ${\rm vac.}$ refers to the vacuum noises due to the sample loss.

Below we consider the \QZ{estimation of the unknown parameter} $\theta=\sum_{j=1}^{K} w^{(j)} \tau_m^{(j)}$ for a specific $m$.

\subsection{Unbiased estimator -- total power constraint}
\QZ{For the total-power-fixed case}, the constraint is on the classical comb amplitude \QZ{$ |A_{m}|^2= 2P_mT/\hbar \Omega_C$. The total power} $P_m$ is then distributed towards the sensors, and we optimize the distribution ratios $\bm c$ here.

For simplicity, \QZ{we assume a uniform signal spectrum} $A_n=A$ \QZ{and a uniform LO spectrum} $B_n^{(j)}=B^{(j)}$. 
The mean value at frequency $n\omega_r + m\Delta \omega_r$ only depends on the BS ratio vector of the first port
\be 
\expval{\hat A_{n,m}^{(j)}} =  \tilde c^{(j)} A
\ee
for $1\le j\le 2K$.

Note that 
\be 
\expval {\hat O_m}= \sum_{j=1}^{2K} v^{(j)}  B^{(j)}   \tau_m^{(j)}  \tilde c^{(j)}  A  
\ee
To get an unbiased estimator $\expval{\hat \theta}=\theta$, we can choose 
\be 
\hat \theta=\hat O_m -  \sum_{j=K+1}^{2K} v^{(j)}  B^{(j)}   \tau_m^{(j)}  \tilde c^{(j)}  A  
\ee
with 
\be 
v^{(j)} B^{(j)}\tilde c^{(j)} A=w^{(j)}
\label{eq:varmin_cond_v_app}
\ee
for $1\le j\le K$. The second term in $\hat \theta$ allows us to optimize $v^{(j)} $ arbitrarily for $j>K$. This will help to maximize quantum squeezing of noise as shown below.

In this case, the estimator variance is
\bal 
{\rm var} \hat \theta &=
\sum_{n=-N}^N \left( |\sum_{j=1}^{2K} v^{(j)}   B^{(j)}  \tau_n^{(j)}  \tilde q^{(j)} |^2 \cdot \frac{1}{G_n} 
+   \sum_{k=1, k\neq K+1}^{2K}  |\sum_{j=1}^{2K} v^{(j)}  B^{(j)}   \tau_n^{(j)}  U^{(j,k)}|^2 
+ |\sum_{j=1}^{2K} v^{(j)}    B^{(j)}   \sqrt{1-|\tau_n^{(j)}|^2}    |^2 \right)\\
\eal 
It can be written in a compact matrix product form 
\bal 
{\rm var} \hat \theta &=
\sum_{n=-N}^N \left(| \bm v\cdot  \bm   B {\bm T}_{n}  \cdot  \tilde  {\bm q} |^2 \cdot \frac{1}{G_n} 
+  \sum_{k=1, k\neq K+1}^{2K}  |\bm v\cdot  \bm   B {\bm T}_{n}  \cdot   \bm u^{(k)}|^2 
+| \bm v\cdot  \bm   B \bm R_{n}   |^2 \right)\\
\label{eq:estimator_var_bothMZIout}
\eal 
\QZ{where the postprocessing weight is $\bm v=[v^{(1)}, \ldots,v^{(2K)} ]^T$, the heterodyne LO amplitude is} $\bm B ={\rm diag} \{B^{(1)}, \ldots, B^{(2K)} \}$, \QZ{the sample transmissivity is} ${\bm T}_{n} = {\rm diag} \{\tau_n^{(1)}, \ldots, \tau_n^{(2K)} \}, \tilde{\bm q}=[q^{(1)}, \ldots,q^{(2K)} ]^T$, \QZ{the sample reflectivity (loss) is} $\bm R_{n} = {\rm diag} \{\sqrt{1-|\tau_n^{(1)}|^2},\ldots, \sqrt{1-|\tau_n^{(2K)}|^2} \}$, \QZ{and the distribution ratio of input vacuum noises towards the sample is} $\bm u^{(k)}=[U^{(1,k)}, \ldots,U^{(2K,k)} ]^T$.

Without loss of generality, we assume $B^{(j)}=1$ since it can be absorbed into $v^{(j)}$.
\QZ{The estimator variance in Eq.~\ref{eq:estimator_var_bothMZIout} is lower bounded by the Cauchy-Schwarz inequality, and the minimum is achieved at} $\tilde q^{(j)}= \frac{v^{(j)}   \tau^{(j)}   }{\sqrt{\sum_{j=1}^{2K} | v^{(j)} \tau^{(j)}  |^2}}$ for $1\le j\le 2K$, and due to the orthogonality of $\bm U$, $  \sum_{j=1}^{2K}v^{(j)} \tau^{(j)}   U^{(j,k)}=0$ for any $k\neq K+1$. Thus given $\bm v$ chosen by Eq.~\eqref{eq:varmin_cond_v_app}, we have $\tilde q^{(j)}=\frac{w^{(j)}\tau^{(j)} /\tilde c^{(j)}  }{\sqrt{\sum_{j=1}^{2K} | w^{(j)} \tau^{(j)} /\tilde c^{(j)}   |^2}}$ for $1\le j\le K$.
We note that $ \tilde c^{(j)}=\tilde c^{(j+K)}=c^{(j)}/\sqrt{2}$ for $1\le j\le K$, $ \tilde q^{(j)}=-\tilde q^{(j+K)}=q^{(j)}/\sqrt{2}$ for $1\le j\le K$, thus the Cauchy-Schwarz minimum condition $\tilde q^{(j)}= \frac{v^{(j)}   \tau^{(j)}   }{\sqrt{\sum_{j=1}^{2K} | v^{(j)} \tau^{(j)}  |^2}}$ requires $v^{(j)}=- v^{(j-K)}\tau^{(j-K)}$ for $K+1\le j \le 2K$. \QZ{Such a choice always satisfies} the unitarity constraint of $\bm U$. 

\QZ{In the neighborhood of the sample lossless limit} $\tau^{(j)}\to 1$, the above optimization gives
\be 
\min {\rm var} \hat \theta = \sum_{n=-N}^N \frac{1}{A^2 G_n}  \left(\sum_{j=1}^{K} |\frac{w^{(j)}}{\tilde c^{(j)}}|^2+ \sum_{j=1}^{K} |\frac{w^{(j)}}{\tilde c^{(K+j)}}|^2\right)   = \sum_{n=-N}^N \frac{2}{A^2 G_n}  \sum_{j=1}^{K} |\frac{w^{(j)}}{\tilde c^{(j)}}|^2 
\label{eq:minvar_totalPower_bothMZIout}
\ee
It agrees with the single-MZI-output \QZ{case in Eq.~\eqref{eq:minvar_totalPower}; thus} one output is sufficient in this weak-absorption limit.

We can further optimize $\tilde c^{(j)}$ by setting $c^{(j)}=\sqrt{|w^{(j)}|}/\sqrt{\sum_{j=1}^{K} |w^{(j)}|}$ \QZ{according to the solution of the Lagrangian multiplier method; then} 
\be 
\min {\rm var} \hat \theta = \sum_{n=-N}^N  \frac{4}{A^2 G_n}   \left(\sum_{j=1}^{K} |w^{(j)}| \right)^2 
\label{eq:minvar_totalPower_bothMZIout_opt}
\ee

\subsection{Unbiased estimator - uniform power constraint}
Plugging in a uniform power allocation vector $c^{(j)}=1/\sqrt{K}$ in Eq.~\eqref{eq:minvar_totalPower_bothMZIout}, we have 
\be 
\min {\rm var} \hat \theta = \sum_{n=-N}^N \frac{4K}{A^2 G_n}  \sum_{j=1}^{K} |w^{(j)}|^2   
\label{eq:minvar_uniform_bothMZIout}
\ee
\QZ{Note that here $A^2$} is the total power over $K$ signal combs. It agrees with the single-MZI-output \QZ{case in Eq.~\eqref{eq:minvar_uniform}; thus} one output is sufficient in this lossless limit.




\section{Distributed division-receiver dual-comb spectroscopy}
\label{app:division}
 
We derive the results of Sec.~\ref{sec:endmatter_division} for an arbitrary
per-sensor absorption spectrum $\{\kappa^{(\ell)}_n\}$, specialising to weak
absorption at the end, and close with the loss sensitivity of the protocol.
 
\subsection{Squeezing structure}
\label{app:div_pairs}
 
Each quantum comb is squeezed at the centre of its own bright lines,
\be
{\rm var}\,\hat P^{A^Q}_n(n)=\frac{1}{G_{A,n}},
\qquad
{\rm var}\,\hat P^{B^Q}_n(0)=\frac{1}{G_{B,n}} ,
\label{eq:div_pairs}
\ee
the self-referred structure of Fig.~5b of Ref.~\cite{shi2025theory}. The readout, however, samples each comb at both centres. At the
other centre the two modes drawn are not \QZ{a squeezed pair,}
\be
{\rm var}\,\hat P^{A^Q}_n(0)=G'_{A,n},
\quad
{\rm var}\,\hat P^{B^Q}_n(n)=G'_{B,n},
\label{eq:div_mismatch}
\ee
\QZ{where }$G'_{n}\equiv\frac12\Bigl(G_{n}+\frac{1}{G_{n}}\Bigr)=1+2N^{\rm amp}_{n}$.
The
classical combs contribute vacuum at either centre,
${\rm var}\,\hat P^{A^C}_n(s)={\rm var}\,\hat P^{B^C}_n(s)=1$.
 
\subsection{Mean readout}
\label{app:div_means}
 
The classical sources carry bright lines and vacuum fluctuations,
\bal
\hat A^{C}(t)&=\frac{1}{\sqrt T}\sum_{n,m}e^{-i(n\omega_r+m\Delta\omega_r)t}
\bigl(A_n\delta_{n,m}+\hat\calA^{C}_{n,m}\bigr),
\nonumber\\
\hat B^{C}(t)&=\frac{1}{\sqrt T}\sum_{n,m}e^{-i(n\omega_r+m\Delta\omega_r)t}
\bigl(B_n\delta_{0,m}+\hat\calB^{C}_{n,m}\bigr),
\label{eq:div_classical}
\eal
\QZ{and $\hat A^{Q},\hat B^{Q}$ have the same form with zero mean.} After the distribution
\eqref{eq:em_dist} and the QCI beamsplitters \eqref{eq:em_qci}, put
$\widetilde A^{(\ell)}_n=c^{(\ell)}A_n/\sqrt2$,
$\widetilde B^{(\ell)}_n=c^{(\ell)}B_n/\sqrt2$ and
\be
\hat\alpha_{\pm,n,m}=\frac{\hat\calA^{C,(\ell)}_{n,m}\pm\hat\calA^{Q,(\ell)}_{n,m}}{\sqrt2},
\qquad
\hat\beta_{\pm,n,m}=\frac{\hat\calB^{C,(\ell)}_{n,m}\pm\hat\calB^{Q,(\ell)}_{n,m}}{\sqrt2}.
\label{eq:div_alphabeta}
\ee
With
$\hat I^{\pm}_{j}=\hat E^{\pm\d}_{j}\hat E^{\pm}_{j}$, we get
\bal
\expval{\hat I^{+}_{1}}&=+\tfrac12\kappa^{(\ell)}_m\widetilde A_m\widetilde B^\star_m,
&\expval{\hat I^{+}_{2}}&=-\tfrac12\widetilde A_m\widetilde B^\star_m,
\nonumber\\
\expval{\hat I^{-}_{1}}&=+\tfrac12\widetilde A_m\widetilde B^\star_m,
&\expval{\hat I^{-}_{2}}&=-\tfrac12\widetilde A_m\widetilde B^\star_m,
\label{eq:div_meanI}
\eal
The readout
is self-calibrating against $\{A_n,B_n\}$ and against $\bm c$, and independent of
the sample phase $\theta_n$.

\subsection{Ratio linearisation and the branch subtraction}
\label{app:div_lin}
 
\QZ{We have}
\be
\calD^{(\ell)}\equiv-\expval{\hat I^{+,(\ell)}_{2}(m\Delta\omega_r)}
=-\expval{\hat I^{-,(\ell)}_{2}(m\Delta\omega_r)}
=\frac{\widetilde A^{(\ell)}_m\widetilde B^{(\ell)\star}_m}{2}.
\label{eq:div_D}
\ee
In the strong-comb limit $A,B\gg1$ the ratios linearise to
\be
\hat r^{(\ell)}_{\rm s}\simeq r^{(\ell)}_{\rm s}
+\frac{\calN^{(\ell)}_{\rm s}}{\calD^{(\ell)}},
\qquad
\hat r^{(\ell)}_{\rm r}\simeq 1+\frac{\calN^{(\ell)}_{\rm r}}{\calD^{(\ell)}},
\label{eq:div_lin}
\ee
with numerators
\bal
\calN^{(\ell)}_{\rm s}&\equiv\Delta\hat I^{+}_{1}(m\Delta\omega_r)
+r^{(\ell)}_{\rm s}\,\Delta\hat I^{+}_{2}(m\Delta\omega_r),
& \\
\calN^{(\ell)}_{\rm r}&\equiv\Delta\hat I^{-}_{1}(m\Delta\omega_r)
+\Delta\hat I^{-}_{2}(m\Delta\omega_r).
\label{eq:div_Ndef}
\eal
Propagating the fluctuations \eqref{eq:div_alphabeta}
through the receiver beamsplitter and the sample, \QZ{and using
\eqref{eq:em_gh}, we obtain}
\bal
\calN^{(\ell)}_{\rm s}
&=\sum_n\Bigl\{g^{(\ell)}_n
\bigl[\widetilde A_n\hat P^{\alpha_+}_n(n)+\widetilde B_n\hat P^{\beta_+}_n(0)\bigr] \\
&+h^{(\ell)}_n
\bigl[\widetilde A_n\hat P^{\beta_+}_n(n)+\widetilde B_n\hat P^{\alpha_+}_n(0)\bigr]\Bigr\}
\nonumber\\
&\quad+\frac12\sum_n\sqrt{2\kappa^{(\ell)}_n\bigl(1-\kappa^{(\ell)}_n\bigr)}
\bigl[\widetilde A_n\hat P^{\calV}_n(n)+\widetilde B_n\hat P^{\calV}_n(0)\bigr],
\label{eq:div_Ns}\\
\calN^{(\ell)}_{\rm r}
&=\sum_n\bigl[\widetilde A_n\hat P^{\alpha_-}_n(n)
+\widetilde B_n\hat P^{\beta_-}_n(0)\bigr].
\eal

Using
$\hat P^{\alpha_\pm}=(\hat P^{A^C}\pm\hat P^{A^Q})/\sqrt2$ and
$\hat P^{\beta_\pm}=(\hat P^{B^C}\pm\hat P^{B^Q})/\sqrt2$, we get
\bal
\calN^{(\ell)}_{\rm s}-\calN^{(\ell)}_{\rm r}
=\frac{1}{\sqrt2}\sum_n\Bigl\{
&\widetilde A_n\Bigl[(g_n\!+\!1)\hat P^{A^Q}_n(n)+(g_n\!-\!1)\hat P^{A^C}_n(n)
+h_n\bigl(\hat P^{B^Q}_n(n)+\hat P^{B^C}_n(n)\bigr)\Bigr]
\nonumber\\
+\;&\widetilde B_n\Bigl[(g_n\!+\!1)\hat P^{B^Q}_n(0)+(g_n\!-\!1)\hat P^{B^C}_n(0)
+h_n\bigl(\hat P^{A^Q}_n(0)+\hat P^{A^C}_n(0)\bigr)\Bigr]\Bigr\}
+\calN_{\calV}.
\label{eq:div_diff}
\eal

\subsection{General-$\kappa$ network variance}
\label{app:div_genkappa}
 
The network observable is built from it by $\hat y^{(\ell)}=\Delta\hat
r^{(\ell)}/2$ and $\hat O_m=\sum_\ell w^{(\ell)}\hat y^{(\ell)}$, that is
\be
\Delta\hat O_m=\frac12\sum_\ell w^{(\ell)}\,
\frac{\calN^{(\ell)}_{\rm s}-\calN^{(\ell)}_{\rm r}}{\calD^{(\ell)}} ,
\ee
and since $\tfrac12 w^{(\ell)}\cdot 2/(c^{(\ell)}B)=p^{(\ell)}_B$ and
$\tfrac12 w^{(\ell)}\cdot 2/(c^{(\ell)}A)=p^{(\ell)}_A$ by \eqref{eq:em_p},
dividing \eqref{eq:div_diff} by $\calD^{(\ell)}$ of \eqref{eq:div_D} gives
\bal
\Delta\hat O_m=\sum_{\ell,n}\Bigl\{
&p^{(\ell)}_B\Bigl[(g_n\!+\!1)\hat P^{A^Q}_n(n)+(g_n\!-\!1)\hat P^{A^C}_n(n)
+h_n\bigl(\hat P^{B^Q}_n(n)+\hat P^{B^C}_n(n)\bigr)\Bigr]
\nonumber\\
+\;&p^{(\ell)}_A\Bigl[(g_n\!+\!1)\hat P^{B^Q}_n(0)+(g_n\!-\!1)\hat P^{B^C}_n(0)
+h_n\bigl(\hat P^{A^Q}_n(0)+\hat P^{A^C}_n(0)\bigr)\Bigr]
+\sqrt{\kappa^{(\ell)}_n\bigl(1-\kappa^{(\ell)}_n\bigr)}
\Bigl[p^{(\ell)}_B\hat P^{\calV}_n(n)+p^{(\ell)}_A\hat P^{\calV}_n(0)\Bigr]\Bigr\}.
\label{eq:div_O}
\eal
 
\QZ{When \eqref{eq:div_O} is regrouped by source, each of the four combs appears} at both
centres, with the weights
\bal
\hat A^{Q}:\;&\bm\pi^{\rm sq}_{A,n}=\bigl(g_n+1\bigr)\bm p_B \;\text{at } s=n,
&&\bm\pi^{\rm mm}_{A,n}=h_n\bm p_A \;\text{at } s=0,
\nonumber\\
\hat B^{Q}:\;&\bm\pi^{\rm sq}_{B,n}=\bigl(g_n+1\bigr)\bm p_A \;\text{at } s=0,
&&\bm\pi^{\rm mm}_{B,n}=h_n\bm p_B \;\text{at } s=n,
\nonumber\\
\hat A^{C}:\;&\bm\pi^{\rm vac}_{A,n}=\bigl(g_n-1\bigr)\bm p_B \;\text{at } s=n,
&&\bm\pi^{\rm mm}_{A,n} \;\text{at } s=0,
\nonumber\\
\hat B^{C}:\;&\bm\pi^{\rm vac}_{B,n}=\bigl(g_n-1\bigr)\bm p_A \;\text{at } s=0,
&&\bm\pi^{\rm mm}_{B,n} \;\text{at } s=n.
\label{eq:div_dressed}
\eal

For a source of paired variance $\zeta_n$
distributed by an array with first column $\bm q$, the vacuum in the unused ports
of \eqref{eq:em_dist} leaves
\be
{\rm cov}\bigl(\hat P^{(\ell)}_n,\hat P^{(\ell')}_n\bigr)
=\delta_{\ell\ell'}+q^{(\ell)}q^{(\ell')}\bigl(\zeta_n-1\bigr).
\label{eq:div_cov}
\ee
\QZ{Contracting with the
weights gives}
\be
{\rm var}\Bigl[\sum_\ell\pi^{(\ell)}\hat P^{(\ell)}_n\Bigr]
=\Psi_n\bigl[\bm\pi,\bm q,\zeta_n\bigr]
\equiv\|\bm\pi\|^2-\bigl(1-\zeta_n\bigr)\bigl(\bm\pi\cdot\bm q\bigr)^2 .
\label{eq:div_Psi}
\ee
Here $\zeta_n=1/G_n$ at a comb's own centre and $\zeta_n=G'_n$ at the other,
by \eqref{eq:div_pairs} and \eqref{eq:div_mismatch}, while $\zeta_n=1$ for any
vacuum, in which case $\Psi_n=\|\bm\pi\|^2$ independently of $\bm q$: an
unsqueezed source acquires no network correlation from being split.
 
Applying \eqref{eq:div_Psi} to the four rows of \eqref{eq:div_dressed} and to the
\QZ{sample vacuum, which is local to each sensor and therefore diagonal, gives}
\be
\boxed{\;
{\rm var}\,\hat O_m=\sum_{n}\Bigl\{
\Psi_n\Bigl[\bm\pi^{\rm sq}_{A,n},\bm q_A,\tfrac{1}{G_{A,n}}\Bigr]
+\Psi_n\Bigl[\bm\pi^{\rm sq}_{B,n},\bm q_B,\tfrac{1}{G_{B,n}}\Bigr]
+\Psi_n\Bigl[\bm\pi^{\rm mm}_{A,n},\bm q_A,G'_{A,n}\Bigr]
+\Psi_n\Bigl[\bm\pi^{\rm mm}_{B,n},\bm q_B,G'_{B,n}\Bigr]
+\calX_n+\calW_n\Bigr\},\;}
\label{eq:div_var_gen}
\ee
the first four terms being the two quantum combs at their two centres and the
last two the vacuum contributions,
\bal
\calX_n&=\bigl\|\bm\pi^{\rm vac}_{A,n}\bigr\|^2+\bigl\|\bm\pi^{\rm vac}_{B,n}\bigr\|^2
+\bigl\|\bm\pi^{\rm mm}_{A,n}\bigr\|^2+\bigl\|\bm\pi^{\rm mm}_{B,n}\bigr\|^2 ,
\\
\calW_n&=\bigl(1+2E_n\bigr)\sum_\ell
\kappa^{(\ell)}_n\bigl(1-\kappa^{(\ell)}_n\bigr)
\Bigl[\bigl|p^{(\ell)}_A\bigr|^2+\bigl|p^{(\ell)}_B\bigr|^2\Bigr] ,
\label{eq:div_W}
\eal
with $E_n$ the thermal occupation of the environment.
 
\subsection{Weak absorption and optimisation}
\label{app:div_weak}
 
At $\kappa^{(\ell)}_n=1$ one has $g_n=1$ and $h_n=0$, so
$\bm\pi^{\rm sq}_{A,n}=2\bm p_B$ and $\bm\pi^{\rm sq}_{B,n}=2\bm p_A$ while every
other dressed weight vanishes together with $\calX_n$ and $\calW_n$;
Eq.~\eqref{eq:div_var_gen} then collapses to Eq.~\eqref{eq:em_var}.
 
The optimisation of Sec.~\ref{sec:em_network} assumed a common classical array
$\bm c$ for the two bright combs. This is not a restriction. Allow independent
arrays $\bm c_A,\bm c_B$, so that $p^{(\ell)}_A=w^{(\ell)}/(c_A^{(\ell)}A)$ and
$p^{(\ell)}_B=w^{(\ell)}/(c_B^{(\ell)}B)$. Minimising each term of
Eq.~\eqref{eq:em_var} over its own splitter as in Eq.~\eqref{eq:em_qopt} gives
\be
{\rm var}\,\hat O^{\rm div}_m\Big|_{\bm q^\star}
=4V\left[\frac{1}{A^2}\sum_\ell\frac{|w^{(\ell)}|^2}{|c_A^{(\ell)}|^2}
+\frac{1}{B^2}\sum_\ell\frac{|w^{(\ell)}|^2}{|c_B^{(\ell)}|^2}\right].
\label{eq:div_afterq}
\ee
The two sums decouple, and each is minimised subject to $\|\bm c_X\|=1$ at
$|c_X^{(\ell)}|^2=|w^{(\ell)}|/\sum_{\ell'}|w^{(\ell')}|$ with common value $W$.
The optimal classical arrays therefore coincide, $\bm c_A=\bm c_B\equiv\bm c$,
which is what makes $\bm p_A\parallel\bm p_B$ and hence $\bm q_A=\bm q_B$ in
Eq.~\eqref{eq:em_optpoint}. With
$\|\bm p_A\|^2+\|\bm p_B\|^2=W(1/A^2+1/B^2)$ and the exposure constraint
$A^2+B^2=4P_m$, Cauchy--Schwarz gives
$1/A^2+1/B^2\ge4/(A^2+B^2)=1/P_m$, with equality at $A=B=\sqrt{2P_m}$.
 
\subsection{Loss sensitivity}
\label{app:div_loss}
 
Away from the lossless limit the $h_n$ terms reinstate
the amplified pairings that limit the protocol.
 
Take a single absorbing line at index $m$ \QZ{against a transparent
background}, $\kappa_n=1$ for $n\neq m$, so that $r_{\rm s}=\kappa$ by
Eq.~\eqref{eq:em_ratio} and, on every background line,
\be
g_n=\frac{1+\kappa}{2},
\qquad
g_n-1=-h_n=\frac{\kappa-1}{2}
\qquad(n\neq m) .
\label{eq:div_ghbg}
\ee
The classical-vacuum and mismatched coefficients are thus \QZ{equal in magnitude, but}
by \eqref{eq:div_Psi} the former is weighted by unity and the latter by
$G'_n\simeq G/2$, so the mismatched term dominates the correction. Evaluating
\eqref{eq:div_var_gen} at the optimum of Sec.~\ref{app:div_weak}, where
$\bm p_A=\bm p_B\equiv\bm p$ and $\bm q=\bm p/\|\bm p\|$, and taking uniform gain
\QZ{$G_n=G$, we find that each of the two squeezed terms} and each of the two mismatched terms
contributes
\be
\underbrace{\frac{(3+\kappa)^2}{4}\,\frac{M}{G}\,\|\bm p\|^2}_{\rm squeezed},
\qquad
\underbrace{\frac{(1-\kappa)^2}{4}\,\frac{MG}{2}\,\|\bm p\|^2}_{\rm mismatched}.
\label{eq:div_twoterms}
\ee
So, 
\be
\frac{\text{mismatched}}{\text{squeezed}}
\simeq\frac{(1-\kappa)^2G^2}{2(3+\kappa)^2}\approx\frac{(1-\kappa)^2G^2}{32} .
\label{eq:div_robust}
\ee
The advantage therefore survives only for shallow
features. The distributed division
receiver thus inherits, sensor by sensor, the entanglement-mismatch pathology of
the single-sensor squeezed division receiver of Ref.~\cite{herman2024squeezed}.


\section{Equivalence between heterodyne and division interferometries
in the weak absorption regime}
\label{Equivalence_Appendix}
 
Write $A^{\rm het}$ for the heterodyne signal comb and $A,B$ for the division
combs of Sec.~\ref{sec:architecture_div}, with lines at sideband index $s=n$ and
$s=0$ respectively; all are taken uniform and one-sided. With
\be
\hat P^{X}_n(s)\equiv\hat\calX^{Q}_{n,s+m}+\hat\calX^{Q\dagger}_{n,s-m},
\qquad
\hat S_X\equiv\sum_n\hat P^{X}_n(s_X),
\label{eq:pair}
\ee
and $s_{A^{\rm het}}=s_B=0, s_A=n $, at $\tau_n\to1$, the two readout noises are, from
\eqref{noise_maintext2_final} and \eqref{eq:em_k1} at $K=1$,
\bal
\Delta\hat d_m&=\hat S_{A^{\rm het}} ,
&
\Delta\bigl(\Delta\hat r_m\bigr)&=4\left[\frac{\hat S_A}{B}+\frac{\hat S_B}{A}\right].
\label{eq:two_noises}
\eal
Let
\be
V\equiv{\rm var}\,\hat S_{A^{\rm het}}={\rm var}\,\hat S_A={\rm var}\,\hat S_B ,
\label{eq:statematch}
\ee
which matches the input covariances, not their preparation. For intra-comb-line squeezing $V=\mathcal S_G\equiv\sum_n1/G_n$, and
$V=M$ classically.
 
Comparing $\hat d_m$ and $\Delta\hat r_m$ as estimators of one parameter through
${\rm var}\,\hat\tau_m={\rm var}(\text{readout})/
|\partial_{\tau_m}\expval{\text{readout}}|^2$ requires the responses
$\partial_{\tau_m}\expval{\hat d_m}=A^{\rm het}/2$ and
$\partial_{\tau_m}\expval{\Delta\hat r_m}=2$.
 
\begin{theorem}[Receiver equivalence]
\label{thm:equivalence}
Under \eqref{eq:statematch}, combs $A$ and $B$ being independent,
\bal
&{\rm var}\,\hat\tau_m^{\rm het}=\frac{4V}{(A^{\rm het})^2},
\qquad
{\rm var}\,\hat\tau_m^{\rm div}
=4V\Bigl(\frac{1}{A^2}+\frac{1}{B^2}\Bigr) \\
&\mathcal R\equiv
\frac{{\rm var}\,\hat\tau_m^{\rm div}}{{\rm var}\,\hat\tau_m^{\rm het}}
=(A^{\rm het})^2\Bigl(\frac{1}{A^2}+\frac{1}{B^2}\Bigr).
\eal
\end{theorem}
 
Heterodyne probes one comb and division the mixture of two, so $\mathcal R$
acquires meaning only at equal exposure, i.e.\ photon number per line delivered to
the sample: $\mathcal P=(A^{\rm het})^2/2$ for heterodyne and $\mathcal P=(A^2+B^2)/4$ for division. Under this constraint ${\rm var}\,\hat\tau_m^{\rm div}$ is
minimised at $A^2=B^2=2\mathcal P$, giving $\mathcal R=2$, attributable to the
division receiver drawing noise from two quantum combs.
 
\subsection*{Distributed heterodyne--division equivalence}
\label{app:eq_dist}
 
The two receivers do not natively estimate the same quantity,
\bal
\expval{\hat d_m^{(\ell)}}&=\tfrac12 A_m^{{\rm het},(\ell)}\bigl(\tau_m^{(\ell)}-1\bigr),
&
\expval{\Delta\hat r_m^{(\ell)}}&
=\bigl(\tau_m^{(\ell)}\bigr)^2-1 ,
\label{eq:two_means}
\eal
so we rescale each sensor to an estimator of $\tau_m^{(\ell)}-1$,
\be
\hat y^{(\ell)}_{\rm het}\equiv\frac{2\,\hat d_m^{(\ell)}}{A_m^{{\rm het},(\ell)}},
\qquad
\hat y^{(\ell)}_{\rm div}\equiv\frac{\Delta\hat r_m^{(\ell)}}{2},
\label{eq:unitresponse}
\ee
the first exactly, the second to leading order since
$\tfrac12[(\tau_m^{(\ell)})^2-1]=(\tau_m^{(\ell)}-1)+O[(1-\tau_m^{(\ell)})^2]$.
For $\hat O_m=\sum_\ell\nu^{(\ell)}\hat y^{(\ell)}$, unbiasedness for $\theta_m$
reads $\nu^{(\ell)}=w^{(\ell)}$ for both receivers; in terms of the original
postprocessing weights this is $\nu^{(\ell)}=v^{(\ell)}A_m^{{\rm het},(\ell)}/2$
for heterodyne and $\nu^{(\ell)}=2v^{(\ell)}$ for division.
 
By \eqref{eq:two_noises} the noises of \eqref{eq:unitresponse} are
$2p_{\rm het}^{(\ell)}\hat S_{A^{\rm het}}^{(\ell)}$ and
$2[p_B^{(\ell)}\hat S_A^{(\ell)}+p_A^{(\ell)}\hat S_B^{(\ell)}]$, with
\be
p_{\rm het}^{(\ell)}=\frac{w^{(\ell)}}{A_m^{{\rm het},(\ell)}},
\qquad
p_A^{(\ell)}=\frac{w^{(\ell)}}{A_m^{(\ell)}},
\qquad
p_B^{(\ell)}=\frac{w^{(\ell)}}{B_m^{(\ell)}}.
\label{eq:net_p}
\ee
Both
\bal
{\rm var}\,\hat O_m^{\rm het}&=F[\bm p_{\rm het},\bm q],
&
{\rm var}\,\hat O_m^{\rm div}&=F[\bm p_B,\bm q_A]+F[\bm p_A,\bm q_B],
\label{eq:net_vars}
\eal
are now optimised by the same route. Minimising over the splitters as in
\eqref{eq:em_qopt} gives $\min_{\bm q}F=4V\|\bm p\|^2$, leaving
$4V\|\bm p_{\rm het}\|^2$ and $4V(\|\bm p_A\|^2+\|\bm p_B\|^2)$. Holding the
per-line exposure $\mathcal P_m$ \QZ{common requires}
\be
\sum_\ell\bigl|A_m^{{\rm het},(\ell)}\bigr|^2=2\mathcal P_m ,
\qquad
\sum_\ell\Bigl[\bigl|A_m^{(\ell)}\bigr|^2+\bigl|B_m^{(\ell)}\bigr|^2\Bigr]
=4\mathcal P_m ,
\label{eq:net_constraint}
\ee
so both pose the convex problem $\min\sum_iu_i^2/x_i$ at fixed $\sum_ix_i=S$,
with $u_i=|w^{(\ell)}|$ and $(K,2\mathcal P_m)$ for heterodyne,
$(2K,4\mathcal P_m)$ for division. Its Lagrange solution
$x_i=S|u_i|/\sum_{i'}|u_{i'}|$ gives $\min=(\sum_i|u_i|)^2/S$, which over the
$2K$ division amplitudes comes out pairwise equal, $A_m^{(\ell)}=B_m^{(\ell)}$,
in agreement with Appendix~\ref{app:div_weak}.
 
\begin{theorem}[Distributed protocol equivalence]
\label{thm:dist_equivalence}
With $W\equiv\bigl(\sum_\ell|w^{(\ell)}|\bigr)^2$, under \eqref{eq:statematch}
and \eqref{eq:net_constraint},
\bal
\min{\rm var}\,\hat O_m^{\rm het}&=\frac{2VW}{\mathcal P_m},
&
\min{\rm var}\,\hat O_m^{\rm div}&=\frac{4VW}{\mathcal P_m},
\label{eq:optvalue}
\eal
both attained at the same
point,
\be
A_m^{{\rm het},(\ell)}=A_m^{(\ell)}=B_m^{(\ell)}
=\sqrt{\frac{2\mathcal P_m|w^{(\ell)}|}{\sum_{\ell'}|w^{(\ell')}|}},
\label{eq:optpoint}
\ee
with $\bm q_A=\bm q_B=\bm q$ and $\bm\nu^{\rm div}=\bm\nu^{\rm het}=\bm w$.
\end{theorem}

\section{Quantum advantage tradeoff without idler}

\label{app:no_idler}

Consider the MZI setup in Fig.~\ref{fig:schematic_MZI_full} \QZ{with $K=2$ samples under a uniform absorption spectrum}, without any idlers $L=0$, and we estimate the weighted \QZ{sum of amplitude transmissivities} $\theta= w_1\tau^{(1)}+w_2\tau^{(2)}$. 

The distribution beamsplitter 
\be 
\bm U=
\bp 
u^{(1,1)} & u^{(1,2)}\\
u^{(2,1)} & u^{(2,2)}
\ep
\ee
\QZ{must be unitary; thus} it has only two relevant degrees of freedom, the \QZ{transmissivity of the first port} $u^{(1,1)}$ and the phase of $u^{(2,1)}$, where $|u^{(1,1)}|^2+ |u^{(2,1)}|^2=1$. The orthogonality requires $u^{(2,2)}=u^{(1,1)}, u^{(1,2)}= -u^{(2,1)}$, up to a global phase.
We inject the classical bright comb at the first port and the quantum squeezed comb \QZ{at the second port; thus, the distribution ratios of the classical comb are} $\tilde {\bm c}=[u^{(1,1)}, u^{(2,1)} ]^T$, $\tilde {\bm q}=[u^{(1,2)}, u^{(2,2)} ]^T=[-u^{(2,1)*}, u^{(1,1)*}]^T$.

We can construct an unbiased estimator 
\be 
\hat \theta= \sum_j w_j \tau^{(j)}
\ee
with $v^{(j)}=\frac{w^{(j)}}{B^{(j)} c^{(j)} A_m}$. 
\QZ{We can write the variance of $\hat \theta$ as }
\bal 
{\rm var} \hat \theta =&
\frac{\left|\tau ^{(1)} \tilde c^{(2)} w^{(1)}/\tilde c^{(1)}-\tau ^{(2)} \tilde c^{(1)} w^{(2)}/\tilde c^{(2)}\right|^2}{G 
   A_m^2}+\frac{\left|\tau ^{(1)} w^{(1)}+\tau ^{(2)} w^{(2)}\right|{}^2}{A_m^2}\\
   &+
   \left(1-|\tau^{(1)}|^2\right) \left| \frac{w^{(1)}}{A_m \tilde c^{(1)}}\right| {}^2 + \left(1-|\tau^{(2)}|^2\right) \left| \frac{w^{(2)}}{A_m \tilde c^{(2)}}\right| {}^2
\eal  
\QZ{In the lossless regime} $\tau^{(1)}, \tau^{(2)}\to 1$
\be 
({\rm var} \hat \theta )^{-1}=
\frac{  A_m^2}{ \left(w^{(1)}+w^{(2)}\right){}^2+ \frac{\left(|\tilde c^{(2)}|^2 w^{(1)}-|\tilde c^{(1)}|^2 w^{(2)}\right){}^2}{ |\tilde c^{(1)}|^2 |\tilde c^{(2)}|^2 G}}
\ee 
We immediately observe that the quantum advantage from $G>1$, \QZ{in comparison with the classical benchmark} $G=1$, only exists when $w^{(1)}$, $w^{(2)}$ have opposite signs, because if they \QZ{have the same sign} $u^{(1,1)}, u^{(2,1)}$ can always be optimized to $u^{(1,1)}=\sqrt{\frac{w_1}{w_1+w_2}}, u^{(2,1)}=\sqrt{\frac{w_2}{w_1+w_2}}$ such that $\left(|u^{(2,1)}|^2 w^{(1)}-|u^{(1,1)}|^2 w^{(2)}\right){}^2=0$. The advantage is maximized when $w^{(1)}=-w^{(2)}$ such that the squeezed noise contribution overwhelms the vacuum noise contribution.

\begin{figure}
    \centering
    \includegraphics[width=0.5\linewidth]{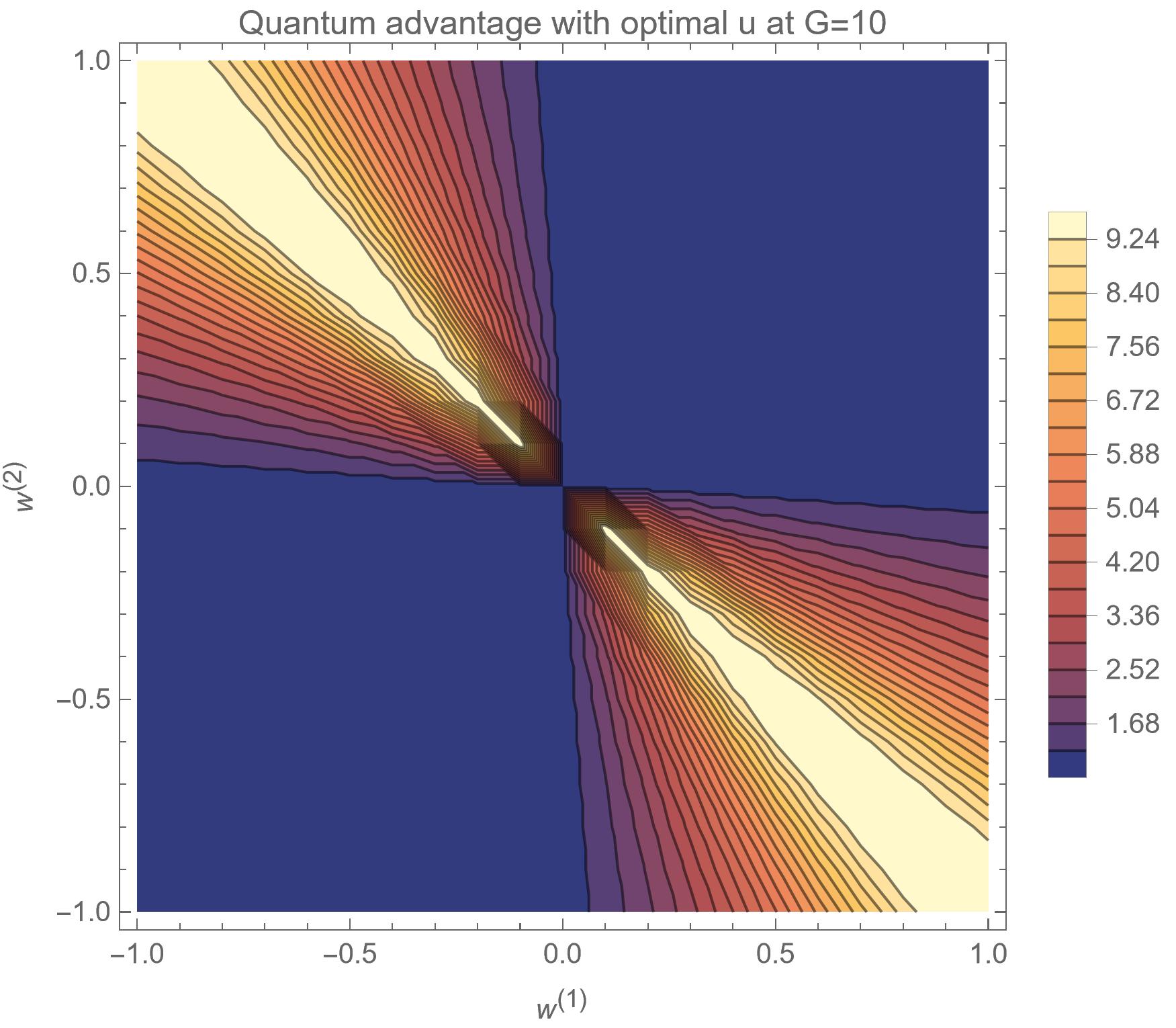}
    \includegraphics[width=0.42\linewidth]{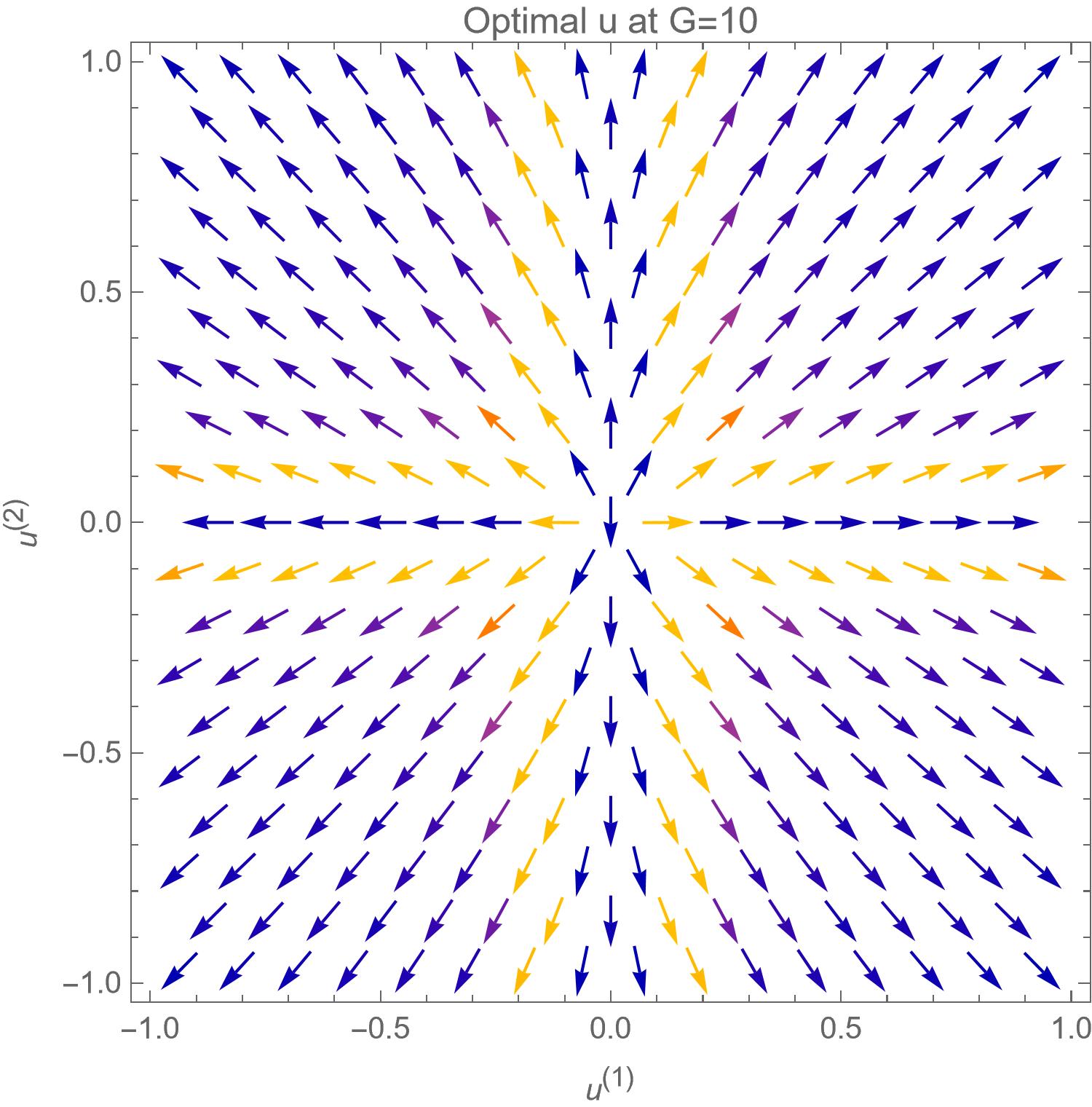}
    \caption{Quantum advantage versus $w^{(1)}, w^{(2)}$. Left: Quantum advantage; Right: the optimum choice of $|A_m^{(1)}|^2/P_m, |A_m^{(2)}|^2/P_m$.}
    \label{fig:Qadv_w1w2}
\end{figure}

\subsection{Necessity
 of the interferometer structure}

One may wonder if one can avoid the interferometer structure, which requires additional modes that do not go through samples. Now we consider the MZI with a quantum comb input and a classical comb input, and we utilize both arms in the MZI  to probe two samples. The full analysis is in Appendix \ref{app:no_idler}. 

In the 2-sample case, we find that quantum advantage diminishes in some target setups when $w_1, w_2$ have the same sign. We numerically evaluate the quantum advantage tradeoff with respect to the target weights $\{w^{(j)}\}_j$ in Fig.~\ref{fig:Qadv_w1w2}.

\end{document}